\documentclass{article}
\usepackage[left=3cm,right=3cm,top=3cm,bottom=3cm]{geometry} % page settings
\usepackage{amsmath} % provides many mathematical environments & tools
\usepackage{authblk}

\RequirePackage[OT1]{fontenc}
\RequirePackage{amsthm,amsmath}
\RequirePackage[numbers]{natbib}
\RequirePackage[colorlinks,citecolor=blue,urlcolor=blue]{hyperref}
\usepackage{amsmath} % provides many mathematical environments & tools

\usepackage{marvosym}
\usepackage{amssymb}
 \usepackage{amsmath}
 \usepackage{natbib}
 \usepackage{float}
 \usepackage{rotating}
 \usepackage{amsmath}
 \usepackage{subfig}
 \usepackage{multirow}
 \usepackage{float}
 \usepackage{mathtools}
 \mathtoolsset{showonlyrefs=true}
\usepackage{xcolor}

\begin{document}

\title{Semi-parametric Bayesian change-point model based on the Dirichlet process%\thanks{Grants or other notes
}


\author[1]{Gianluca Mastrantonio}

\affil[1]{Department of Mathematical Science, { Politecnico di Torino} \&  International Association for Research Seismic Precursors (iAReSP)}
\date{   }

\maketitle

\begin{abstract}

In this work we introduce a   semi-parametric Bayesian  change-point model, defining its time dynamic as   a  latent Markov process based on  the Dirichlet process. We treat the number of change point as a random variable and  we estimate it during model fitting. Posterior inference is carried out using a Markov chain Monte Carlo algorithm based on a marginalized version of the proposed model.

The model is illustrated using  simulated examples and two real datasets, namely the coal-mining disasters, that is a widely used dataset for illustrative purpose, and a dataset of indoor radon recordings.
With the simulated examples we show that  the model is able to recover the parameters and  number of change points, and we compare our results with the ones  of the-state-of-the-art models, showing a clear improvement in terms  of  change points identification.
The results obtained on  the coal-mining disasters  and radon data are coherent with previous literature.

\end{abstract}

\section{Introduction}

{A change-point model is a mixture-type model used to infer changes in a time series subjected to random shifts in its characteristics/features.} This means that the  data can   be broken down into segments and  each segment follows a  statistical model  with different parameters.  The time when a segment ends  is called  \emph{change point} and the segment is often referred to as \emph{regime} or \emph{state}. The inference based on  change-point models  focuses on two  major issues: i) the estimate  of number  and locations  of the change-points; ii) the choice of  the best statistical model for each segment.

The change-point literature, starting from \cite{Quandt1958} and \cite{chernoff1964}, is by now fairly extensive in both frequentist   \citep{BHATTACHARYA1987183,Hawkins2001}  and Bayesian framework  \citep{Carlin1992,Giordani2008,Chaturvedi2015}.
In the former model estimation can be difficult  since the likelihood function becomes rapidly intractable as the number of change points increases \citep[for a discussion see][]{elliott2003}. On the other hand,  in the more recently developed Bayesian models, the estimation procedures, generally based on Markov chain Monte Carlo (MCMC) algorithms, is always feasible,  raising  attention to this modelling approach. Among the existing Bayesian models,  the most commonly used is the one proposed by \cite{Chib1998} \cite[see for example][]{Pastor2001,Chang2004,koop2004,ko2015}.    
%one of them is the one proposed by \cite{Chib1998}.

In \cite{Chib1998}  a time series is modelled introducing a latent realization of a discrete time series, that denotes the  regime membership, with  temporal evolution ruled by a first order Markov process. The change-point model is then obtained assuming a  transition matrix constrained so that regimes are visited in a non-reversible sequence; the model can then be seen as a constrained hidden Markov model (HMM)  \citep[for an extensive introduction on the HMM, see][]{zucchini2009b}.
% and this similarity facilitates the definition and implementation of the Markov chain Monte Carlo (MCMC) algorithm, {built} to estimate model parameters. 

In \cite{Chib1998}, the Bayes factor is used to asses the number of segments through an  off-line procedure.  
Informational criteria, such as the  Bayes factor, AIC and BIC, has been  criticized \citep{Dziak}   since they often suggest different models and it is not always clear which one is the most trustworthy. In  a Bayesian setting,   we can replace the  information criteria with a fully probabilistic on-line model choice,  that can be based  on the reversible-jumps  \citep{GREEN1995} or Dirichlet process (DP)  \citep{Ferguson1973}.\\
The reversible-jump is a Markov chain Monte Carlo (MCMC)  algorithm that 
simulates from  posterior distributions defined on  spaces of varying dimensions
and it can be used to perform model choice. Its implementation  requires  a mapping function between model parameters  that is not always straightforward to define and it has a great impact on the ability of the MCMC to explore the target distribution   \citep{Brooks2003}. 
On the other hand, the DP can be used as a prior for an infinite set of parameters, it allows 
to perform model choice  in mixture-based models \citep{teh2010}  and, generally, it leads to  MCMC algorithms  straightforward to implement.

In this work we propose a semi-parametric extension of \cite{Chib1998} based on the DP,  which address issue i) in a fully probabilistic setting, allowing an on-line model choice, while the second issue is left to future developments and considered out of the scope of this work.

{Prior to this work,} \cite{Kozumi2000} and \cite{ko2015}  dealt with \cite{Chib1998} extensions   DP based. Both  of them have   flaws that make their use  problematic. 
In \cite{Kozumi2000}, as also noted by \cite{ko2015}, {no  temporal evolution in the  latent allocation dynamic is considered,  a regime can always be revisited and the model reduces to a  mixture. In   \cite{ko2015}  there is not a clear and rigorous formalization of the underlying DP, there are  incorrect computations of some full conditionals and   the proposed  MCMC algorithm  updates  the latent allocations  in a way that easily leads to the identification of the wrong number of regimes (more details on these issues are given in  the Appendix).} The model of \cite{ko2015} is close to our proposal and then, together with the one of   \cite{Chib1998},  are considered as our main competitor.

In this paper we explain how to use the DP to build a semi-parametric extension of \cite{Chib1998}, giving  a  rigorous formalization of the entire procedure. Semi-parametric HMMs based on the DP has been previously proposed, see for example by \cite{Teh2006} and  \cite{fox2011}, but here, due to the peculiar transition matrix, these approaches cannot be used. We propose  to use the  DP  to obtain  countably infinite   distributions, each one with only two possible outcomes and where the probabilities of the outcomes are  related to the stick-breaking weights \citep{sethuraman:stick}.
 This approach {allows us} to treat the number of segments as  random  and to estimate it during  model fitting.  {Our specification of the  model  induces  issues in the regime labeling that are   solved  by  using a  collapsed Gibbs sampler   \citep{Liu1994} that marginalizes over the DP weights; the sampling algorithm is partially based on an birth and death  MCMC.

    Our proposal  is applied to simulated datasets and two real ones.
    The formers are used to  show how the  proposed MCMC 
    algorithm is able to recover model parameters,  number and positions of the   change points. Our results are compared with the ones
    of  \cite{Chib1998} and \cite{ko2015} and we show that  a great improvement in terms of  change points identification is achieved. The models are then  applied to one of the most used test-dataset in change-point studies, i.e.  the coal-mining disasters data \cite[see for example][]{JARRETT1979,Carlin1992}. The results we obtain are consistent  with the one of \cite{Chib1998}, but, under our model, we are  able to give a measure of uncertainly on the number of latent change points. In the last example  a  time series of Italian indoor radon measurement \citep{Nicol2015} is  analyzed.
    {Radon emissions are characterized by a non-stationary temporal pattern with  periodic components \citep{Baykut2010} at different time scales \citep{barbosa2010} and
        changes in mean, variability and trend.     
        Radon concentration is  considered a possible earthquake precursor \citep{Woith2015} since have been observed that, prior to strong earthquakes, abrupt changes in the time series characteristics occur. The segmentation of radon data  is a first step to try to understand its connection with geodynamic activity.
        To the best of our knowledge, in the literature have never been proposed a model-based method to segment a radon time series while, for example,
        wavelet transformation \citep{Barbosa2007} and
        %     ,  empirical mode decomposition \cite{Baykut2010}
        testing procedures \citep{barbosa2010}  have been exploited.
        We show that our model identifies  reasonable change points and with   sojourn time in a regime of about  a day,  that was   also  observed in previous studies \citep[see for example][]{barbosa2010}, proving  that  
        change-point models can be used to infer changes in a radon time series. }

    The paper is organized as follows. In Section \ref{sec:DP} we introduce the DP.  In Section \ref{sec:chib} we formalize the model of Chib and in Section \ref{sec:model} we show our proposal. The MCMC algorithm is shown in Section \ref{sec:mcmc} while Section \ref{sec:ex} contains the simulated and real data  examples. The paper ends with a discussion in Section \ref{sec:disc}. In the Appendix  
    we 
    {highlight} what we believe are the problematic aspects and unclear points of the model and MCMC implementation proposed by \cite{ko2015}.

    \section{The semi-parametric change-point  model} \label{sec:cp}

Before the model specification, we  introduce the DP.

    \subsection{The Dirichlet process} \label{sec:DP}

    The DP   is a stochastic process defined over a measurable space $(\Theta,\mathcal{B})$ \citep{Ferguson1973}  and it is  a random probability
    measure on a space of distribution functions, i.e. a drawn from a DP is a random  discrete distribution, it
    depends on a \emph{scaling parameter} $\beta>0$ and a \emph{base distribution} $H$ over $\Theta$; the density of $H$  will be indicated with $h(\cdot)$. By definition   $G$ is DP distributed with parameters $(\beta,H)$, i.e. $G|\beta,H \sim DP(\beta,H) $, if for any finite  partition $\{A_k\}_{k =1}^K$ of $\Theta$ such that $\cup_{k=1}^K A_k \equiv \Theta$ and $A_k \cap A_{k^{\prime}}= \{\emptyset\}$ if $ k\neq k^{\prime}$, 
    we  have    
    \begin{align} \label{eq:g}
    &(G(A_1),G(A_2), \dots G(A_K))^{\prime}|\beta,H  \sim \\ &\phantom{s}Dir(\beta H(A_1), \beta H(A_2),\dots , \beta H(A_K)),
    \end{align}
    where $Dir(\cdot,\cdot, \dots,\cdot)$ indicates the Dirichlet distribution.
    Since 
    \begin{align} \label{eq:beta}
    &(G(A),1-G(A))^{\prime}|\beta,H \sim\\& \phantom{s} Dir(\beta H(A), \beta(1-H(A)) )\equiv B(\beta H(A),  \beta(1-H(A))),
    \end{align}
    where $B(\cdot,\cdot)$ is the beta distribution, mean and variance of $G(A)$ can be easily computed:
    \begin{equation} \label{eq:meanvar}
    E(G(A)) = H(A) , \quad Var(G(A))= \frac{H(A)(1-H(A))}{\beta+1}.
    \end{equation}
    From \eqref{eq:meanvar} we see that 
    $H$ is   the expected shape of $G$ while $\beta$  controls the degree  of variability.

    \cite{sethuraman:stick}  gives an  explicit representation of $G$, that is called the \emph{stick-breaking process} or  \emph{stick-breaking representation}; If  
    \begin{equation} \label{eq:discG}
    G= \sum_{k \in \mathbb{N}} \tau_k \delta_{\boldsymbol{\theta}_k},
    \end{equation}
    is DP distributed, then 
    \begin{align}
    \pi_{k} &\sim B(1, \beta),\label{eq:piw}\\
    \tau_k& = \pi_{k}\prod_{l=1}^{k-1}(1-\pi_l)\label{eq:pi},\\
    \boldsymbol{\theta}_k &\sim H \label{eq:piH},
    \end{align}
    where $\delta_{\cdot}$ is a point mass function, $\{\tau_k\}_{k \in \mathbb{N}}$ is the set of weights and  $\{\boldsymbol{\theta}_k\}_{k \in \mathbf{N}}$    the set of \emph{atoms} of the DP.
    Notice that $\tau_k>0$, $\sum_{k \in \mathbb{N}}\tau_k = 1 $ and   G is then  a discrete distribution. Sets
    $\{\tau_k\}_{k \in \mathbb{N}}$ and $\{\pi_k\}_{k \in \mathbb{N}}$ are often  written as $\{\tau_{\boldsymbol{\theta}_k}\}_{k \in \mathbb{N}}$ and $\{\pi_{\boldsymbol{\theta}_k}\}_{k \in \mathbb{N}}$ to stress their connection with the  DP atoms  $\{\boldsymbol{\theta}_k\}_{k \in \mathbb{N}}$.   
    For computational purposes  \cite[see for example][]{neal2000,gelfand2005} a drawn from a  DP is frequently parametrized using  $\{\tau_k, \boldsymbol{\theta}_k\}_{k=1}^{\infty}$.
    % and, even if it is  not  customary, 
    %the one-to-one relation between  sets $\{\tau_k\}_{k \in \mathbb{N}}$ and  $\{\pi_k\}_{k \in \mathbb{N}}$, makes possible to express a DP drawn using $\{\pi_k, \boldsymbol{\theta}_k\}_{k=1}^{\infty}$, i.e.  
    %$$G= \sum_{k \in \mathbb{N}} \tau_k \delta_{\boldsymbol{\theta}_k} =  \sum_{k \in \mathbb{N}} \pi_{k}\prod_{l=1}^{k-1}(1-\pi_l) \delta_{\boldsymbol{\theta}_k} . 
    %$$ 
    %%This representation will be useful in the building of our model, see Section \ref{sec:model}. 
    %
    %%\textcolor{red}{FIN QUI 30/06}

    The discrete nature of $G$, with its countably infinite  atoms and weights, makes the use of the DP convenient to extend semi-parametrically mixture-based models, where the  couples atom-weight are potential sets of parameters (the atoms) and mixture probabilities (the weights);
    details can be found in  \cite{antoniak1974},  \cite{Mac1998},  \cite{teh2010} or \cite{fox2011}. 
    
    %Given a finite sample of $T$ observations, the number of observable  unique mixture components  is bounded, i.e. it cannot exceed $T$, and \textcolor{red}{their number is used as an estimate of the in-sample mixture dimensionality.  Qui mi perdo il soggetto, intendi dire che il numero di componenti \`e usato come stima della in-sample dimensionality? Ma perch\`e dirlo qui? a cosa serve?}

    %As we state in the introduction, the model of \cite{Chib1998} can be seen as a HMM, that is a particular mixture model,  and,  then the DP can be uses to define a semi-parametric extension.

    \subsection{The model of \cite{Chib1998}} \label{sec:chib}
    
    {
        
        In this section we  introduce the hierarchical model of  \cite{Chib1998}.
        %   For clarity of notation we will often use  symbols defined in  Section \ref{sec:DP}. \\
        Let $\mathbf{y}=\{y_t\}_{t=1}^T$ be an observed time series. At the first level  the conditional density of   $y_{t}|\{y_j\}_{j=1}^{t-1}$\footnote{We assume $\{ y_j \}_{j=1}^0 \equiv  \{  \emptyset \}$.}   is assumed to depend  on a vector of parameters $\boldsymbol{\theta}_{s_t} \in \Theta$,  indexed by a discrete latent random variable $s_t \in \{1,2,\dots,K^*\}$ that indicates the regime membership, i.e. if $s_t=k$  then $y_t$ belongs to the $k^{th}$ regime; notice that $\boldsymbol{\theta}_{s_t} \equiv \boldsymbol{\theta}_{k}$ if $s_{t}=k$.
        At   the second level  $\{s_t\}_{t=1}^T$  is  a Markov process, with starting point $s_1=1$,  ruled by a $K^* \times K^*$ constrained \emph{one-step ahead} transition matrix
        \begin{equation}
        P= \left(
        \begin{array}{ccccccc}
        \pi_{1} & 1-\pi_1 & 0 & 0 & \cdots & 0 \\
        0 & \pi_2 & 1-\pi_2 &0& \cdots & 0\\
        0 & 0 & \pi_3 & 1-\pi_3 & \cdots & 0\\
        0 & 0 & 0 & \pi_4 & \cdots & 0 \\
        \cdots & \cdots & \cdots & \cdots &  \cdots\\
        0 & 0 & 0 & 0 & \cdots & 1
        \end{array}
        \right).\label{eq:tra}
        \end{equation}
        Since the lower diagonal elements of $P$ are zeros,   a regime  left cannot be visited again.}

    Let  $f(\cdot)$ indicate a density function and  $I(\cdot,\cdot)$  the indicator function,  we can then write 
    \begin{align}
     &f(s_t|s_{t-1}=k, \{ \pi_k \}_{k=1 }^{K^*} ) =\\ & \phantom{s} \pi_k I(s_t,k)+(1-\pi_k)  I(s_t,k+1),\, t=2,\dots , T,\\
    &s_1=1,
    \end{align}
    where it is assumed that $s_t \in\{k,k+1\}$ if  $s_{t-1}=k$.

    \cite{Chib1998} assumes  beta distributions with the same set of parameters for all the elements of $P$ and then, letting  $H$  be  a prior distribution, the model can be written as 
    \begin{align}
    &f(\mathbf{y}|\{\boldsymbol{\theta}_k  \}_{k=1}^{K^*},\{s_t\}_{t=1}^T) = \prod_{t=1}^T \prod_{k =1}^{K^*} f(y_t|\{ y_j \}_{j=1}^{t-1},\boldsymbol{\theta}_{k})^{I(s_t,k)},\\
&   f(s_t|s_{t-1}=k, \{ \pi_k \}_{k=1 }^{K^*} ) =       \\
&\phantom{s}\pi_k I(s_t,k)+(1-\pi_k)  I(s_t,k+1), \, t=2,\dots ,T, \label{eq:qqq}\\
&   s_1=1,\label{eq:qqq1}\\
    & \pi_{k} \sim B(\alpha,\beta),\, k=1,\dots K^*,\\
&   \boldsymbol{\theta}_k  \sim H,\, k=1,\dots K^*.
    \end{align}
    The model described above can be seen as an HMM with constrained transition matrix.
    The number of   
    rows of   $P$, that is equal to the number of regimes, must be set \emph{a priori} (see equation \eqref{eq:tra}) and an off-line procedure is needed to assess the value of $K^*$. With our proposal we are going to extend the model of    \cite{Chib1998} allowing an on-line model choice.

    \subsection{The semi-parametric extension} \label{sec:model}

    Our extension starts with the introduction of  an equivalent   specification of the  model of Chib that  is obtained by substituting the latent process  $\{s_t\}_{t=1}^T $
    %\textcolor{red}{$s_t$ ma nel paragrafo precedente non era $s_t$? secondo me devi cambiare sopra $s_t$ in $s_t$, anche per essere congruente con l'introduzione} 
    with  $\{\boldsymbol{\psi}_t \in \Theta\}_{t=1}^T$,  assuming the following time evolution:
    \begin{align}
    &f( \boldsymbol{\psi}_t|\boldsymbol{\psi}_{t-1}= \boldsymbol{\theta}_k, \{ \pi_k \}_{k=1 }^{K^*} ) \sim \\ &\phantom{s} \pi_k I(\boldsymbol{\psi}_t,\boldsymbol{\theta}_k)+(1-\pi_k)  I(\boldsymbol{\psi}_t,\boldsymbol{\theta}_{k+1}),  \, t=2,\dots , T,\label{eq:sss}\\
&   \boldsymbol{\psi}_1  = \boldsymbol{\theta}_1.\label{eq:sss1}
    \end{align}
    assuming $\boldsymbol{\psi}_t \in \{\boldsymbol{\theta}_k, \boldsymbol{\theta}_{k+1} \}$ if $\boldsymbol{\psi}_{t-1}= \boldsymbol{\theta}_k$.

    Notice that  equations \eqref{eq:qqq} and \eqref{eq:qqq1} are equivalent to  equations \eqref{eq:sss} and \eqref{eq:sss1} since $f(\boldsymbol{\psi}_t=\boldsymbol{\theta}_k|\boldsymbol{\psi}_{t-1}=\boldsymbol{\theta}_{k^{\prime}}) = f(s_t=k|s_{t-1}=k^{\prime})$ and 
    $\boldsymbol{\psi}_t=\boldsymbol{\theta}_k$   if and only if  $s_t=k$.

    %\textcolor{red}{ Tutta questa frase non la capisco bene, si prende il valore asintotico del numero di componenti ($T$)? oppure prendi semplicemente $K^*=\infty$? (direi questa soluzione, nel caso cambierei la frase: Semi-parametric extensions for  mixture-based model  are generally built by taking  the number of components $K^*= \infty $ , and assuming a DP prior for the probability structure of the latent allocation process  } \\
    
    {Semi-parametric extensions for  mixture-based models  are generally defined by taking   $K^*\rightarrow \infty $ and assuming a DP based prior for the probability structure of   $\{\boldsymbol{\psi}_t\}_{t=1}^T$ \citep[see for example][]{Escobar1995,Teh2006,Johnson2013}.  
        Here we propose the following. First  notice that each row of  $P_{\theta}$  sums to 1, i.e. is a vector of probabilities, { with only two non-zero values}. We assume  $G= \sum_{k=1}^{\infty}\tau_k \delta_{\theta_k} \sim DP(\beta,H)$ and 
        we  define distributions $G_{\theta_k}$'s, with $k \in \mathbb{N}$, as follows: 
        \begin{equation} \label{eq:Gthetaw}
        G_{\theta_{k}} = \frac{\tau_{k}}{1-\sum_{l=1}^{k-1}\tau_{l}}\delta_{\theta_k}+\left(    1-\frac{\tau_{k}}{1-\sum_{l=1}^{k-1}\tau_{l}}  \right)\delta_{\theta_{k+1}},\, k \in \mathbb{N}.
        \end{equation}
        {In our model  $G_{\theta_k}$ is used as  distribution for  the  $k^{th}$ row of $P_{\theta}$.  Notice that  $\frac{\tau_{k}}{1-\sum_{l=1}^{k-1}\tau_l}$ is equal to the beta distributed weight $\pi_k$  (see equation \eqref{eq:pi}),  and then \eqref{eq:Gthetaw}  can be written equivalently  as}
        \begin{equation} \label{eq:Gtheta}
        G_{\theta_{k}} = \pi_{k}\delta_{\theta_k}+(1-\pi_{k})\delta_{\theta_{k+1}},\, k \in \mathbb{N} ,
        \end{equation}
        where by definition, see Section \ref{sec:DP}, 
        \begin{align}
        \pi_k & \sim B(1,\beta), \label{eq:qw}\\
        \theta_k & \sim H. \label{eq:qw2}
        \end{align}
        We have then  distributions based on the DP, one for each row of the infinite-dimensional transition matrix $P_{\theta}$. Notice that the atoms in the regimes are tied by construction, i.e. the atom of $\left[P_{\theta}\right]_{i,i+1}$ is equal to the one of $\left[P_{\theta}\right]_{i+1,i+1}$. We can then write   
        \begin{align}
        &\boldsymbol{\psi}_t|\boldsymbol{\psi}_{t-1}= \theta_k, \{\boldsymbol{\theta}_k, \pi_k\}_{k \in \mathbb{N}}  \sim G_{\theta_{k}} \label{eq:dd} 
        , \, t=2,\dots ,T,\\
    &   \boldsymbol{\psi}_1  = \boldsymbol{\theta}_1.
        \end{align}
        %The temporal evolution of $\boldsymbol{\psi}_t$ is defined using the  couples $(\pi_k,\boldsymbol{\theta}_k)^{\prime}$s, with $k \in \mathbb{N}$. 
        { The model is then }
        %
        %As stated in Section \ref{sec:DP}, we can  express a draw from a DP in terms of $\{\boldsymbol{\theta}_k\}_{k\in \mathbb{N}}$ and the stick-breaking     weights $\{\pi_k\}_{k \in \mathbb{N}}$, that are Beta distributed  (see Section \ref{sec:DP} and equation \eqref{eq:piw}). The sets $\{\boldsymbol{\theta}_k\}_{k\in \mathbb{N}}$ and $\{\pi_k\}_{k \in \mathbb{N}}$ are then used to  define the  distributions:
        %
        %Notice that $G_{\theta_{k}}$ is equal to the non zero elements of the  $k^{th}$ row of $P_{\theta}$.
        %Then we define the  time dynamics of  the latent allocation process as 
        %\begin{align}
        %\boldsymbol{\psi}_t|\boldsymbol{\psi}_{t-1}= \theta_k, \{\boldsymbol{\theta}_k, \pi_k\}_{k \in \mathbb{N}}  &\sim G_{\theta_{k}} \label{eq:dd} 
        %, \, t=2,\dots ,T,\\
        %\boldsymbol{\psi}_t & = \boldsymbol{\theta}_1.
        %\end{align}
        %The transition matrix that rules the process \eqref{eq:dd} is precisely the one in \eqref{eq:tra33}.  \\Finally, the  semi-parametric extension of Chib's model is the following: 
        \begin{align}
    &   f(\mathbf{y}|\{\boldsymbol{\psi}_t  \}_{t=1}^T) = \prod_{t=1}^T  f(y_t|\{ y_j \}_{j=1}^{t-1},\boldsymbol{\psi}_{t}),\\
    &   \boldsymbol{\psi}_t|\boldsymbol{\psi}_{t-1}= \boldsymbol{\theta}_k, \{\boldsymbol{\theta}_k, \pi_k\}_{k \in \mathbb{N}}  \sim G_{\theta_{k}} 
        ,\, t=2,\dots ,T,\\
    &   \boldsymbol{\psi}_1  = \boldsymbol{\theta}_1,\\
    &   \pi_{k}|\beta  \sim B(1, \beta), \, k \in \mathbb{N} ,\\
    &   \boldsymbol{\theta}_k|H  \sim H, \, k \in \mathbb{N},
        \end{align}
        or, introducing the discrete time series $\{s_t\}_{t=1}^T$,  it can be  equivalently stated as 
        \begin{align}
    &   f(\mathbf{y}|\{\boldsymbol{\theta}_k  \}_{k\in \mathbb{N}},\{s_t\}_{t=1}^T) = \prod_{t=1}^T \prod_{k \in \mathbb{N}} f(y_t|\{ y_j \}_{j=1}^{t-1},\boldsymbol{\theta}_{k})^{I(s_t,k)},\\
    &   f(  s_t|s_{t-1}=k, \{ \pi_k \}_{k \in \mathbb{N}} )  =  \\ & \phantom{s} \pi_k I(s_t,k)+(1-\pi_k)  I(s_t,k+1), \, t=2,\dots ,T,\\
    &   s_1=1,\\
    &   \pi_{k}|\beta  \sim B(1, \beta),\, k \in \mathbb{N},\\
    &   \boldsymbol{\theta}_k|H  \sim H,\, k \in \mathbb{N}.
        \end{align}
        This model is an infinite-dimensional extension of the one shown at the end of Section \ref{sec:chib}.\\ 
        As in the standard DP based mixture models, the number  $K$ of unique values  that $\boldsymbol{\psi}$ (or $s$) assumes,   is used as an  estimate of the number of  segments of the observed time series. 
        Notice that  $H$ acts as the  prior distribution of $\boldsymbol{\theta}_k$.
        %This last model will be  specification is used  to define the Bayesian estimation algorithm. 

        \section{The MCMC algorithm} \label{sec:mcmc}

        From equation \eqref{eq:Gtheta}  and  matrix $P_{\theta}$ we see that regimes are visited in increasing order, e.g. after regime $k$, regime  $k+1$ is visited and  this can  produce  an
        inefficient MCMC algorithm. 
        Then,
        to avoid the problem, we marginalized over the vector of DP weights. This strategy is often adopted  \citep[see for example][]{neal2000,Teh2006,Blei2010}  since the resulting process defines  a prior over a partition of the data    that no more depends on the labels.ì  
        Let $n_{i}^{j:j^{\prime}}= \sum_{t=j}^{j^{\prime}-1} \delta(s_t,i)\delta(s_{t+1},i)$, that is the number of self-transitions in the $i^{th}$ regime  between time $j$ and $j^{\prime}$. After   marginalization {we obtain the following  for the dynamic of } $s_t$:
        \begin{align} 
        &f(s_{t}=i|s_{t-1}=k,s_{t-1},\dots , s_{1},\beta)= \\ 
    & \phantom{}    \begin{cases}
        \frac{n_{k}^{1:(t-1)}+1}{n_{k}^{1:(t-1)}+1+\beta} & \mbox{if } i=k,\\ 
        \frac{\beta}{n_{k}^{1:(t-1)}+1+\beta} & \mbox{if } 
        %j \neq s_h,h=1,2,\dots , t
        i=k+1,\\
        \end{cases}t=2,\dots , T,
        \label{eq:s_t}\\
    &   s_1     =1.
        \end{align}
        We want to remark that now regimes are visited in increasing order only  to simplify the  notation, but any regimes re-labeling are equivalent.   \\
        {The conditional distribution of $s_{t}$ depends on the count $n_{k}^{1:(t-1)}$ and  parameter $\beta$ and the process $s_t$  is no more Markovian.} 
        {The probability of $s_t=k|s_{t-1}=k,s_{t-1},\dots , s_{1},\beta$, i.e. $s_t$  assumes the same value of  $s_{t-1}$, increases with $n_{k}^{1:(t-1)}$ meaning that, if at time $t$  an observation is allocated to the previously observed  regime $k$, at time $t+1$ the probability to belong to the  same regime   increases; i.e. the process has the \emph{self reinforcement}  property \citep{pemantle2007}. Parameter $\beta$ can be interpreted noticing that when  there is only one observation in the $k^{th}$ regime, i.e. $n_{k}^{1:(t-1)}=0$, the odd to move to a new regime at time $t+1$ is $\beta$.}

        {The model  is then}
        \begin{align}
    &   f(\mathbf{y}|\{\boldsymbol{\theta}_k  \}_{k\in \mathbb{N}},\{s_t\}_{t=1}^T) = \prod_{t=1}^T \prod_{k \in \mathbb{N}} f(y_t|\{y_j\}_{j=1}^{t-1},\boldsymbol{\theta}_{k})^{I(s_t,k)},\\
    &   f(s_{t}=i|s_{t-1}=k,s_{t-2},\dots , s_{1},\beta)= \\ & \phantom{s}
        \begin{cases}
    &   \frac{n_{k}^{1:(t-1)}+1}{n_{k}^{1:(t-1)}+1+\beta}\, \mbox{if } i=k,\\ 
    &   \frac{\beta}{n_{k}^{1:(t-1)}+1+\beta}\,  \mbox{if } 
        %j \neq s_h,h=1,2,\dots , t
        i=k+1,\\
        \end{cases} t=2,\dots,T,\\
        &s_1  = 1,\\
    &   \boldsymbol{\theta}_k|H  \sim H, \, k \in \mathbb{N}.
        \end{align}

        Under this setting the MCMC  updates of  $\beta$ and $\boldsymbol{\theta}_k$  are simple and we show, in the next paragraphs, how to implement them. The update of  $\{s_t\}_{t=1}^T$ will be discussed in more details since it needs to be more carefully implemented to obtain and efficient algorithm.
        
        \paragraph{The update of  $\beta$}
        
        Let $f(\beta)$ be a prior distribution, then
        the full conditional of $\beta$ is   proportional to 
        $
        f(s_1,\dots,s_T|\beta) f(\beta). 
        $
        Using \eqref{eq:s_t} we can find that  
        \begin{align}
    &   f(s_1,\dots,s_T|\beta)=\\ & \phantom{s} \left[ \prod_{i=1}^{K-1}     \frac{ \beta \prod_{j=0}^{n_{i}^{1:T}-1}(1+j) } {  \prod_{j=0}^{n_{i}^{1:T}}(1+\beta+j) }  \right]   \frac{  \prod_{j=0}^{n_{K}^{1:T}-1}(1+j) } {  \prod_{j=0}^{n_{K}^{1:T}-1}(1+\beta+j) } , \label{eq:prod}
        \end{align}
        and using   relation $a(a+1)\dots (a+m-1)= \frac{\Gamma(a+m)}{\Gamma(a)}$, \eqref{eq:prod} can be expressed as
        \begin{equation} \label{eq:S}
        %\left[ \prod_{i=1}^{K-1}  \frac{\beta \Gamma(\beta+1) \Gamma(n_{i}^{1:T}+1  )}{\Gamma(n_{i}^{1:T}+1+\beta+1  )} \right] \frac{\beta \Gamma(\beta) \Gamma(n_{K,K}^{1:T}+1  )}{\Gamma(n_{K,K}^{1:T}+1+\beta  )}=
        %\end{equation}
        %\begin{equation}\label{eq:fst2}
        \beta^{K-1}\prod_{i=1}^{K}  \frac{ \Gamma(\beta+1) \Gamma(n_{i}^{1:T}+1  )}{\Gamma(n_{i}^{1:T}+1+\beta+1 -I(i,K) )}.
        \end{equation}
        The  full conditional of $\beta$ is then
        \begin{equation} \label{eq:fullbeta}
        \beta^{K-1}\prod_{i=1}^{K}  \frac{ \Gamma(\beta+1) \Gamma(n_{i}^{1:T}+1  )}{\Gamma(n_{i}^{1:T}+1+\beta+1 -I(i,K) )} f(\beta).
        \end{equation}

    {To the best of our knowledge, there is not a prior distribution $f(\beta)$ that let us  express \eqref{eq:fullbeta}  in a closed form from which sampling is easy and a sample of $\beta$  must be draw using a  Metropolis-Hastings step. }

        \paragraph{The update of $\boldsymbol{\theta}_k$}

        The full conditional of  $\boldsymbol{\theta}_k$  is proportional to 
        \begin{equation} \label{eq:fulltheta}
        \prod_{t=1}^{T}     f(y_{t}|\{y_j\}_{j=1}^{t-1}, \boldsymbol{\theta}_{k}) ^{I(s_t,k)} h(\boldsymbol{\theta}_k),
        \end{equation}
        The functional form of \eqref{eq:fulltheta} depends on how we specify $f(y_{t}|\{ y_j \}_{j=1}^{t-1}, \boldsymbol{\theta}_{k})$ and  $H$. As an example,  if  $f(y_{t}|\{ y_j \}_{j=1}^{t-1}, \boldsymbol{\theta}_{k})\equiv f(y_{t}| \boldsymbol{\theta}_{k})$, with  $Y_t |\boldsymbol{\theta}_{k} \sim N(\mu_k,\sigma_k^2)$ and   $\boldsymbol{\theta}_k= \{ \mu_k,\sigma_k^2 \}$, then if  $H$ is the product of a normal distribution over $\mu_k$ and   inverse gamma over $\sigma_k^2$,  likelihood and prior  are conjugate and the full conditional is normal-inverse gamma \citep{Gelman2003}. %With  non conjugate prior and likelihood, the algorithm of \cite{neal2000} can be used to obtain posterior samples.  

        \subsection{The update of $\{s_t\}_{t=1}^T$}

        It is generally preferable to update  jointly as many random variables as possible \citep{Robert2005}.   
        Unfortunately, we are unable to find a way  to sample from the joint full condition of $\{s_t\}_{t=1}^T$ and then a different approach must be used. 
        A simple solution is the univariate update of each component  $s_t$ but, experimenting with simulated data, we notice that {this } leads to unsatisfactory  results in terms of MCMC chain mixing since, for example,       redundant states  with similar $\boldsymbol{\theta}_k$'s are created and the distribution of $K$ is generally {entirely} concentrated on a  single value. We solved the aforementioned problems  by  combining the univariate update with  other   updates:
        \begin{itemize}
            \item  the split update (or birth move) -  we propose a new change point  at time $t$;
            %\begin{enumerate}[.a]
            %   \item starting form $s_1$ to $s_T$;
            %   \item  starting form $s_T$ to $s_1$;
            %\end{enumerate}
            \item  the merge update (or death  move) - we propose to merge consecutive  regimes.
        \end{itemize}
%       It is customary to simulate the elements of a time series   sequentially, starting from time 1 to time $T$, but 
%   tests on simulated datasets showed  that MCMC mixing is improved if, choosing randomly, the univariate and split updates   are performed   starting from the first to the last time or from the last to the first. \\ Notice that, to simplify  notation, after each MCMC step regimes are relabelled so they  satisfy   $s_1=1$ and $s_{t} \in \{s_{t-1}, s_{t-1}+1  \}$.  
    {At each MCMC iteration only one of them  is performed, choosing   randomly with assign probabilities. }    We  assume that before the MCMC update of  $s_t$ is performed, it have value $k$ and,  to simplify  notation, after each MCMC step regimes are relabelled so to  $s_1=1$ and $s_{t} \in \{s_{t-1}, s_{t-1}+1  \}$.
        \paragraph{The single-component update}
        Let $n_{i}^{-t}= n_{i}^{1:(t-1)} +n_{i}^{(t+1):T}$, i.e. the number of self transition in the $i^{th}$ regime without taking into account  transitions that involve  $s_{{t}}$, and let  $*$ indicates a new regime.
        We have to sample  $s_t$ only if $s_t\neq s_{t-1}$ or $s_t\neq s_{t+1}$, otherwise $s_t = s_{t+1}=s_{t-1}$ with probability 1, then  
        \begin{itemize}
            \item with probability proportional to $$\frac{\beta}{\beta+1}f(y_{t}|\{ y_{j}\}_{j=1}^{t-1}, \boldsymbol{\theta}_{*})$$
            $s_t=*$; 
            \item if $t \neq 1$, $s_t$ can be equal to $s_{t-1}$ with probability proportional to 
            \begin{equation}
            \frac{n_{s_{t-1}}^{-t}+1}{n_{s_{t-1}}^{-t}+1+\beta+1}  f(y_{t}|\{ y_{j}\}_{j=1}^{t-1},\boldsymbol{\theta}_{s_{t-1}});
            \end{equation}
            \item if $t \neq T$, $s_t$ can be equal to $s_{t+1}$ with probability proportional to
            \begin{equation}
            \frac{n_{s_{t+1}}^{-t}+1}{n_{s_{t+1}}^{-t}+1+\beta+1-I(s_{t+1},K)}   f(y_{t}|\{ y_{j}\}_{j=1}^{t-1}, \boldsymbol{\theta}_{s_{t+1}});
            \end{equation}
            \item 
            if  $n_{k}^{1:T}=0$,  then  $s_{t-1} \neq k \neq s_{t+1}$, and  $s_t$ can be equal to $k$ with full conditional 
            \begin{equation}
            \propto \frac{\beta}{\beta+1}f(y_{t}|\{ y_{j}\}_{j=1}^{t-1}, \boldsymbol{\theta}_{k}).
            \end{equation}
        \end{itemize}
        %Indeed $K$ increases if $s_t = *$ and $n_{i^*,i^*}^{-t}\neq0$, while it decreases when $n_{i^*,i^*}^{-t} = 0$ and the sample of $s_{t}$ is equal to  $s_{t-1}$ or $s_{t+1}$. 

        \paragraph{The split update}
        Let $S_{-} = \{s_{t'}:s_{t'}=k, t^{\prime}<t \}$ and $S^{+} = \{s_{t'}:s_{t'}=k, t^{\prime}\geq t \}$, let $n_{S_{-}}$ and $n_{S^{+}}$ be the number of self transitions in the two subsets and let
        \begin{equation}
        \gamma_{c}(n) = \frac{\Gamma(\beta+1)\Gamma( n+1  )}{\Gamma(n+1+\beta+1 -c)},
        \end{equation}
        then 
        \begin{itemize}
            \item  $s_t=k$ for all  $s_t \in S_{-}\cup S^+$ with probability  
            \begin{equation}
            \propto 
            \gamma_{I(k,K)}(n_{k}^{1:T})
            \prod\limits_{t: s_t  \in S_{-}\cup S^+ }  f(y_{t}|\{ y_{j}\}_{j=1}^{t-1}, \boldsymbol{\theta}_{k}) ;
            \end{equation}
            \item $s_t = *$ for all $s_t \in S_{-}$ and   $s_t = k$ for all $s_t \in S^+    $ with probability  
            \begin{align}
            &\propto\beta 
            \gamma_{0}(n_{S_{-}})
            \gamma_{I(k,K)}(n_{S^{+}})\times  \\
            & \phantom{s}\prod\limits_{t: s_t  \in S_{-} }  f(y_{t}|\{ y_{j}\}_{j=1}^{t-1}, \boldsymbol{\theta}_{*}) 
            \prod\limits_{t: s_t   S^+ }  f(y_{t}|\{ y_{j}\}_{j=1}^{t-1}, \boldsymbol{\theta}_{k});
            \end{align}   
            \item $s_t = k$  for all $s_t \in S_{-}$  and  $s_t = *$ for all $s_t \in S^+$ with probability 
            \begin{align}
            &\propto\beta 
            \gamma_{0}(n_{S_{-}})
            \gamma_{I(k,K)}(n_{S^{+}})   \times \\
        &\phantom{s}    \prod\limits_{t: s_t  \in S_{-} }  f(y_{t}|\{ y_{j}\}_{j=1}^{t-1}, \boldsymbol{\theta}_{k}) 
            \prod\limits_{t: s_t   S^+ }  f(y_{t}|\{ y_{j}\}_{j=1}^{t-1}, \boldsymbol{\theta}_{*}).
            \end{align}
        \end{itemize}

        \paragraph{Merge update}
        Let $S_{j}=\{ s_t: s_t=j\}$, then,  for $k=1,\dots , K$:
        \begin{itemize}
            \item $s_t = *$ for all $s_t$ in $S_{k}$ with probability
            \begin{align}
        &   \beta  \gamma_{0}({n}_{k-1,k-1}^{1:T})^{1-I(k,1)}  \gamma_0({n}_{k,k}^{1:T}) \gamma_{I(k,K)}({n}_{k+1,k+1}^{1:T})^{1-I(k,K)}\times  \\
        &\phantom{s}    \prod \limits_{t: s_t  \in S_k }  f(y_{t}|\{ y_{j}\}_{j=1}^{t-1}, \boldsymbol{\theta}_{*});
            \end{align}
            \item if $k \neq 1$, then $s_t= k-1$ for all $s_t$ in $S_{k}$ 
            with probability
            \begin{align}
        &   \gamma_{0}({n}_{k-1,k-1}^{1:T}+{n}_{k,k}^{1:T}+1) \gamma_{I(k,K)}({n}_{k+1,k+1}^{1:T})^{1-I(k,K)}  \times \\
        &\phantom{s}    \prod \limits_{t: s_t  \in S_k }  f(y_{t}|\{ y_{j}\}_{j=1}^{t-1}, \boldsymbol{\theta}_{k-1});
            \end{align}
            \item if $k \neq K$, then $s_t= k+1$ for all $s_t$ in $S_{k}$
            with probability
            \begin{align}
        &   \gamma_{0}({n}_{k-1,k-1}^{1:T})^{1-I(k,1)} \gamma_{I(k,K)}({n}_{k,k}^{1:T}+{n}_{k+1,k+1}^{1:T}+1)\times  \\
        &\phantom{s}    \prod \limits_{t: s_t  \in S_k }  f(y_{t}|\{ y_{j}\}_{j=1}^{t-1}, \boldsymbol{\theta}_{k+1});
            \end{align}
            \item $s_t = k$ for all $s_t$ in $S_{k}$  with probability
            \begin{align}
        &   \beta  \gamma_{0}({n}_{k-1,k-1}^{1:T})^{1-I(k,1)}  \gamma_0({n}_{k,k}^{1:T})  \times \\
        &\phantom{s}\gamma_{I(k,K)}({n}_{k+1,k+1}^{1:T})^{1-I(k,K)} \prod \limits_{t: s_t  \in S_k }  f(y_{t}|\{ y_{j}\}_{j=1}^{t-1}, \boldsymbol{\theta}_{k}).
            \end{align}
        \end{itemize}

        MCMC mixing is improved if, choosing randomly, the univariate and split updates   are performed   starting from the first to the last time or from the last to the first. 
        . 
        
        \section{Examples} \label{sec:ex}

        In this section we compare the results of our model with the ones of   \cite{ko2015} and \cite{Chib1998} on simulated datasets and real ones.
        Using  simulated datasets we test the ability of the models in recovering the right number of latent regimes and parameters, then the models are estimated on a standard change-point problem, that is the number of coal-mining disasters,  and  on the radon data.
        %
        %Our results are   compared with the ones obtained using the model and MCMC  algorithm of \cite{ko2015} and \cite{Chib1998}. 
        We implement the model of  \cite{Chib1998}, introduced in  Section \ref{sec:chib}, assuming prior \eqref{eq:qw}  for the transition probabilities and the same priors over $\beta$ and  likelihood parameters as the one of our proposal;  model choice is performed using  BIC.
        The model of \cite{ko2015}, with respect to our approach,  starts from a different specification  of the latent process $s_t$ (for details see the Appendix) with
        \begin{align}
    &   f(s_{t}=i|s_{t-1}=k,s_1,\dots , s_{t-1},\beta, \alpha)= \\
    &\phantom{s}    \begin{cases}
        \frac{n_{k}^{1:(t-1)}+\alpha}{n_{k}^{1:(t-1)}+\alpha+\beta} & \mbox{if } i=k,\\ 
        \frac{\beta}{n_{k}^{1:(t-1)}+\alpha+\beta} & \mbox{if } 
        %j \neq s_h,h=1,2,\dots , t
        i=k+1,\\
        \end{cases}
        \label{eq:s_t2s}
        \end{align}
        that reduces to our specification if $\alpha=1$, see equation  \eqref{eq:s_t}. Then, in this section, the model of \cite{ko2015} is implemented using the MCMC algorithm that the authors proposed,  assuming $\alpha=1$ and under   the same priors over $\beta$ and likelihood parameters of our proposal. In all the examples posterior inference is carried out using  130000 iterations, burnin 80000 and thin 10, using 5000 posterior values for inferential purposes, with an half normal prior for $\beta$ with variance parameter $\sigma_{\beta}^2$.
    {We chose between
        the single component, split and  merge updates with probabilities respectively equal to 0.5, 0.25 and 0.25, while with probability 0.5 we perform the univariate or split update starting from the first to the last time. We indicate our proposal as  Model1, the model of \cite{Chib1996} as Model2 and the one of \cite{ko2015} as Model3. }

        The \emph{R} \citep{rcran} source codes that can be used to replicate the results of the simulated  and  coal-mining disasters  examples are available online in a GitHub repository\footnote{\url{https://github.com/GianlucaMastrantonio/Change-Point} }
while, due to  a confidentiality issue, only  the \emph{R} functions used to analyze the radon example are available.

        \subsection{Simulated data} \label{sec:simex}
        \begin{figure*}[t]
            \centering
            {\includegraphics[scale=0.45]{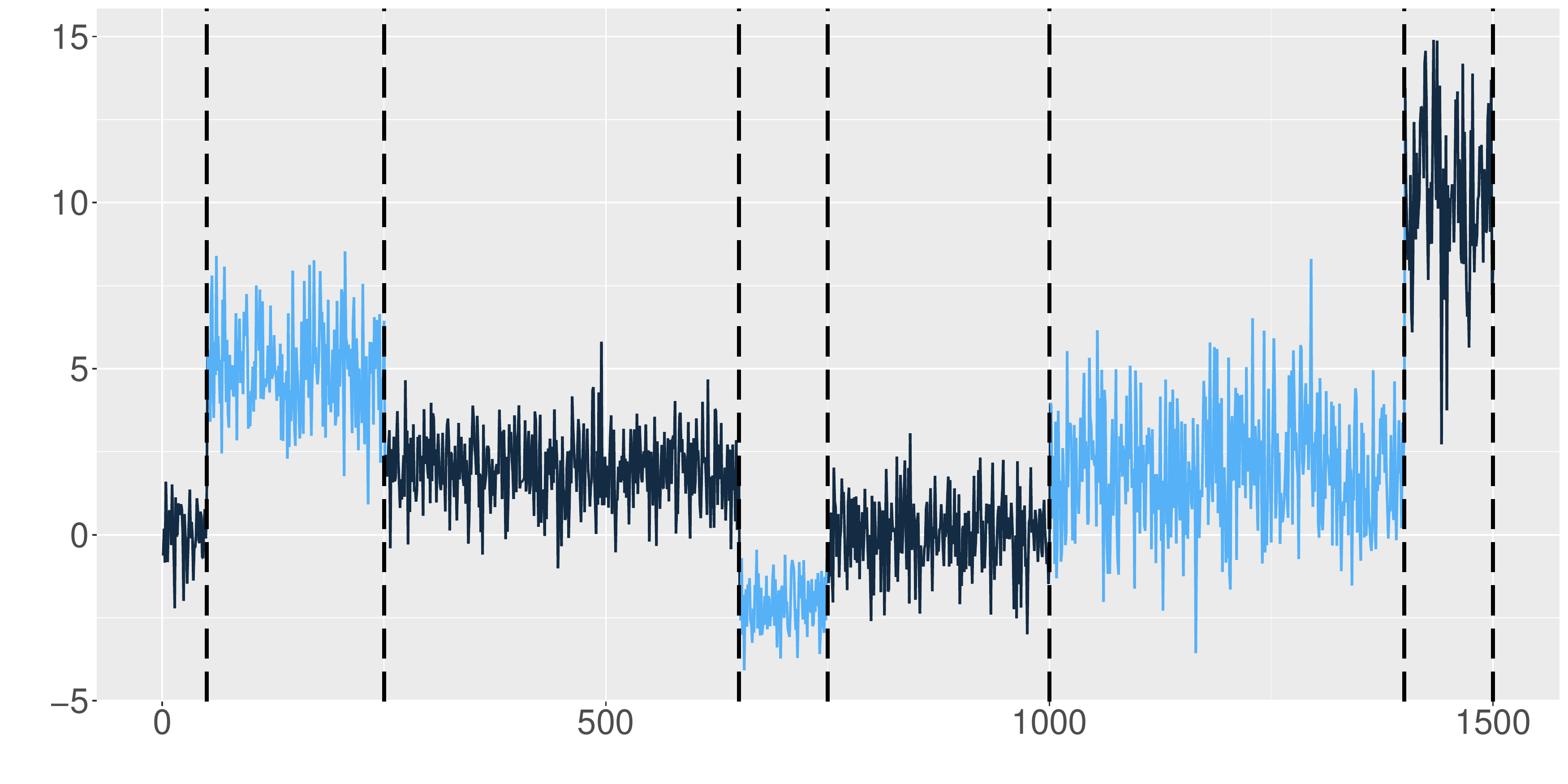}}
            \caption{Simulated example -  first scheme: one of the simulated times series. The vertical dashed lines separate the regimes.}\label{fig:exy1}
        \end{figure*}
        
        \begin{figure*}[t]
            \centering
            {\includegraphics[scale=0.45]{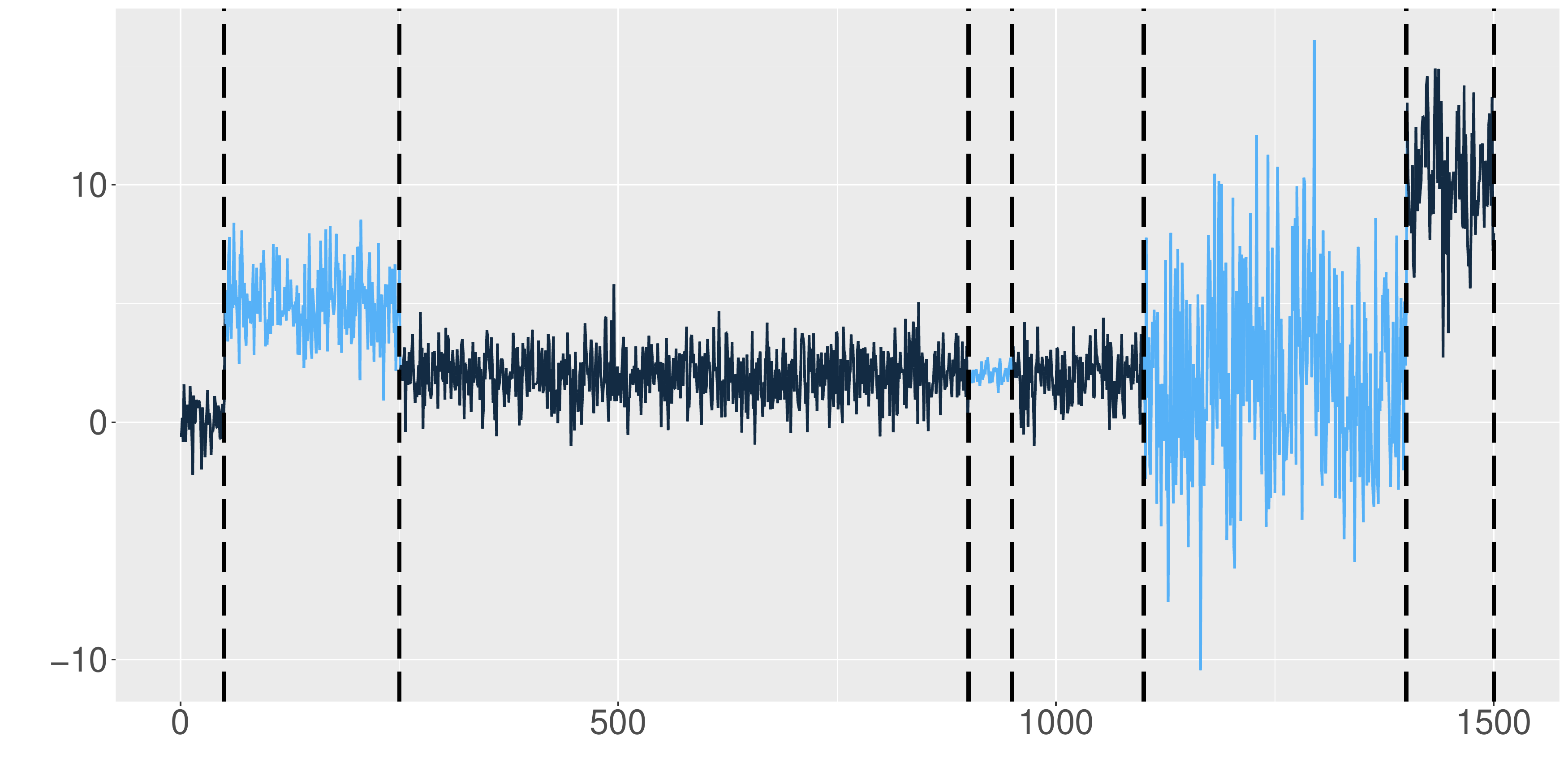}}
            \caption{Simulated example -  second scheme: one of the simulated times series. The vertical dashed lines separate the regimes.}\label{fig:exy12}
        \end{figure*}

        \begin{figure*}[t]
            \centering
            {\includegraphics[scale=0.45]{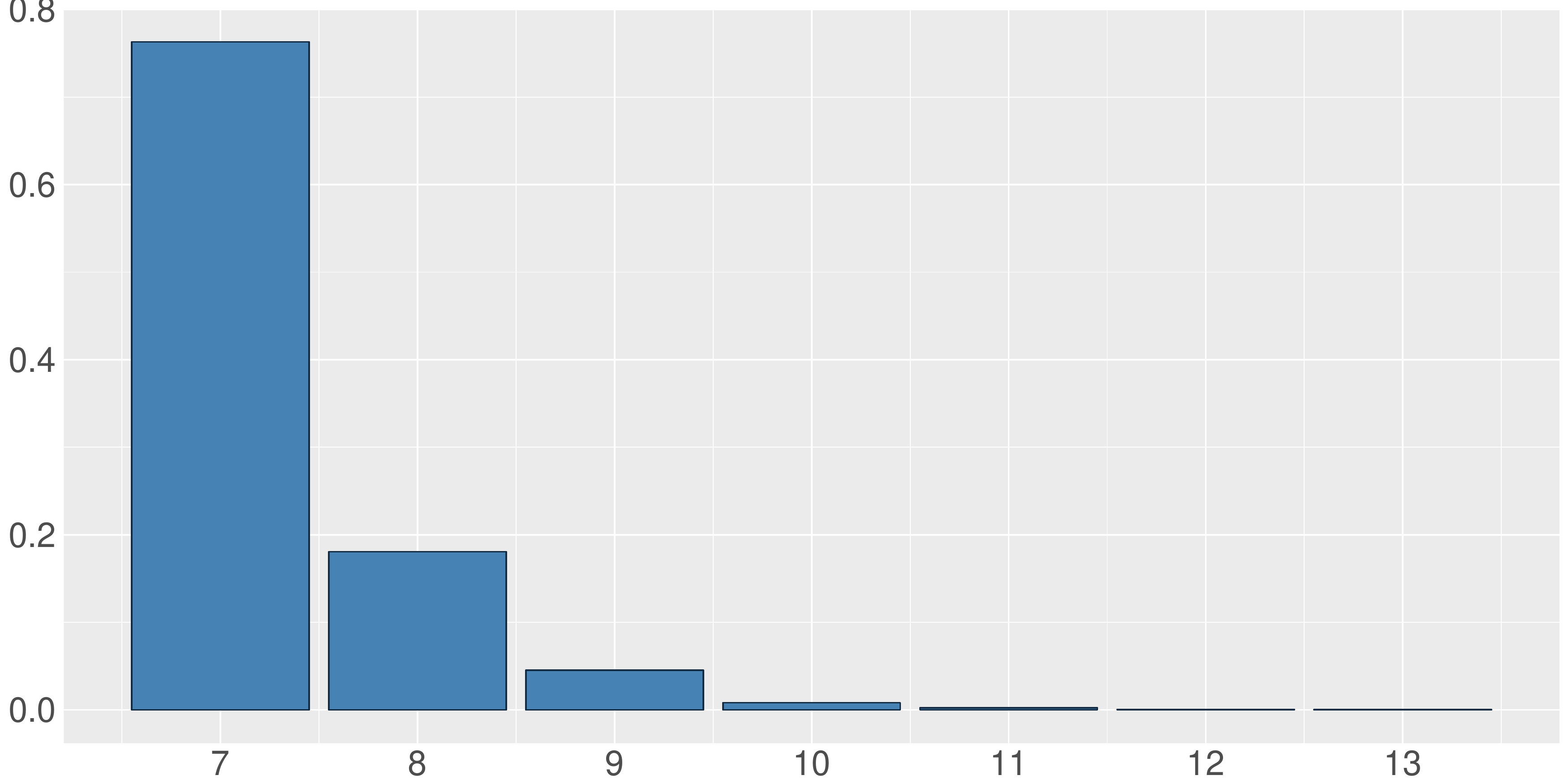}}
            \caption{Simulated example -  first scheme: posterior distribution of $K$.}\label{fig:K1}
        \end{figure*}

        \begin{figure*}[t]
            \centering
            {\includegraphics[scale=0.45]{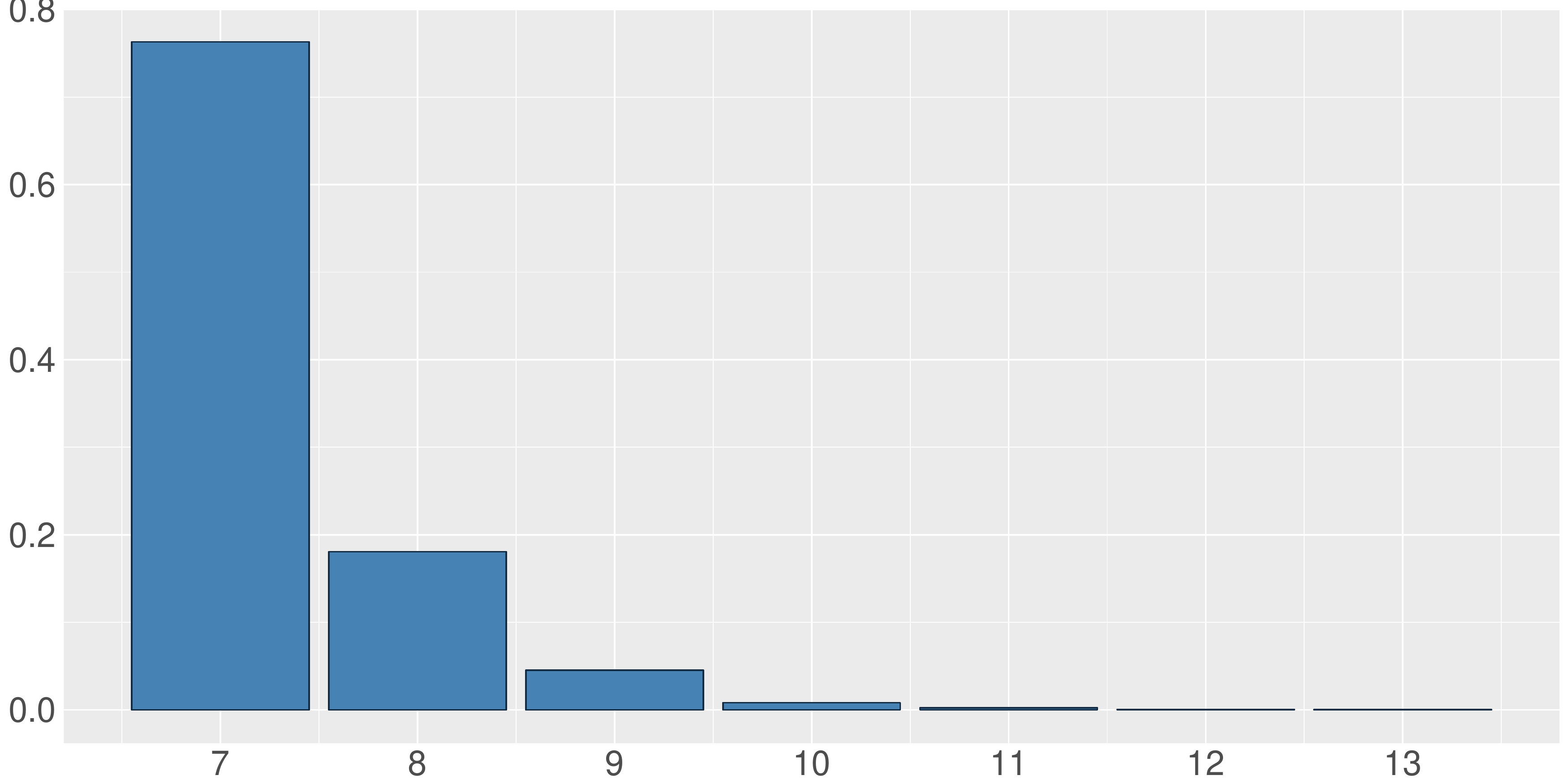}}
            \caption{Simulated example -  second scheme: posterior distribution of $K$.}\label{fig:K2}
        \end{figure*}

        \begin{table}[t]
            \centering
            \begin{tabular}{cc|cccccccccc}
                \hline \hline
                &&$\hat{\mu}_k$& $\hat{\sigma}_k^2$ &   $\hat{\xi}_k$\\
                && (CI) & (CI)& (CI)\\
                \hline
                &1 & 0.1 &0.711 & 50 \\
                &  & [-0.139  0.341] & [0.491 1.092]& [50 50]\\
                &2 &5.005 &1.968 & 250 \\
                &  & [4.806 5.206] & [1.636 2.403]& [250 251]\\
                &3& 1.985&1.081 & 650\\
                &  & [1.881 2.088] & [0.945 1.245]& [650 650]\\
                &4 &-2.084 &0.628 &749 \\
                &  & [-2.249 -1.919] & [0.481 0.842]& [747   753]\\
                &5 &-0.002 &1.132 & 1000\\
                &  & [-0.134  0.132] & [0.954 1.359]& [998  1000]\\
                &6 &1.956 &3.017 &1400 \\
                &  & [1.791 2.125] & [2.644 3.448]& [1400 1400]\\
                &7 &10.184 & 5.361&  \\
                &  & [ 9.714 10.640] & [4.126 7.088]& \\
                \hline \hline
            \end{tabular}
            \caption{Simulated example -  first scheme - Model1:  posterior means  $ (\,\hat{ }\, )$  and credible intervals (CI) of  $\mu_k$, $\sigma_k^2$ and  $\xi_k$ computed using the subset of posterior samples that has $K=7$. } \label{tab:ex}
        \end{table}
        
        \begin{table}[t]
            \centering
            \begin{tabular}{cc|cccccccccc}
                \hline \hline
                &&$\hat{\mu}_k$& $\hat{\sigma}_k^2$ &   $\hat{\xi}_k$\\
                && (CI) & (CI)& (CI)\\
                \hline
                &1 &0.1 & 0.713&  50\\
                &  & [-0.136  0.342] & [0.494 1.079]&[50 50]  \\
                &2 &5.003 &1.978 & 250 \\
                &  & [4.809 5.200] & [1.629 2.416]& [250 251] \\
                &3 & 1.986& 1.083& 650 \\
                &  & [1.882 2.088] & [0.948 1.249]& [650 650] \\
                &4 & -2.084& 0.633&749  \\
                &  & [-2.243 -1.921] & [0.485 0.851]& [747 753] \\
                &5 & 0.001&1.129 &  1000\\
                &  & [-0.135  0.135] & [0.953 1.360]& [998  1000 ] \\
                &6 &1.957 &3.018 &1400  \\
                &  & [1.786 2.133] & [2.633 3.483]& [1400  1400] \\
                &7 &10.18 &5.349 &  \\
                &  & [9.725 10.632] & [4.125 7.167]& \\
                \hline \hline
            \end{tabular}
            \caption{Simulated example -  first scheme - Model2:  posterior means  $ (\,\hat{ }\, )$  and credible intervals (CI) of  $\mu_k$, $\sigma_k^2$ and  $\xi_k$. } \label{tab:ex2}
        \end{table}

        \begin{table}[t]
            \centering
            \begin{tabular}{cc|cccccccccc}
                \hline \hline
                &&$\hat{\mu}_k$& $\hat{\sigma}_k^2$ &   $\hat{\xi}_k$\\
                && CI & CI& CI\\
                \hline
                &1 &0.101 &0.715 & 50\\
                &  & [-0.138  0.336] & [0.493 1.095]& [50    50]\\
                &2 & 5.008&1.974 &250\\
                &  & [4.816 5.199] & [1.627 2.420]& [250 251]\\
                &3 & 1.972& 1.123&900\\
                &  & [1.891 2.050] & [1.014 1.257]& [900 903]\\
                &4& 2.08& 0.163&957\\
                &  & [1.966 2.188] & [0.114 0.245]& [950   959]\\
                &5 & 1.935& 1.118&1101\\
                &  & [1.758 2.107] & [0.896 1.432]& [1098  1101]\\
                &6 & 1.887 & 14.833&1400\\
                &  & [1.448 2.332] & [12.638 17.546]& [1399  1400]\\
                &7 & 10.171& 5.376&  \\
                &  & [9.720 10.625] & [4.088 7.161]& \\
                \hline \hline
            \end{tabular}
            \caption{Simulated example -  second scheme - Model1:  posterior means  $ (\,\hat{ }\, )$  and credible intervals (CI) of  $\mu_k$, $\sigma_k^2$ and  $\xi_k$ computed using the subset of posterior samples that has $K=7$. } \label{tab:ex3}
        \end{table}
        
        \begin{table}[t]
            \centering
            \begin{tabular}{cc|cccccccccc}
                \hline \hline
                &&$\hat{\mu}_k$& $\hat{\sigma}_k^2$ &   $\hat{\xi}_k$\\
                && CI & CI& CI\\
                \hline
                &1 &0.099 &0.712 &50\\
                &  & [-0.138  0.340] & [0.496 1.071]& [50    50]  \\
                &2& 5& 1.978&250\\
                &  & [4.808 5.199] & [1.637 2.419]& [250   251] \\
                &3 & 1.973& 1.123&900\\
                &  & [1.895 2.058] & [1.008 1.257]& [899   905] \\
                &4 &2.077 & 0.162&956\\
                &  & [1.963 2.190] & [0.111 0.238]& [947   959] \\
                &5 & 1.937&1.115  &1101 \\
                &  & [1.767 2.113] & [0.89  1.42]& [1096  1101]\\
                &6 &1.897 & 14.808&1400\\
                &  & [1.458 2.341] & [12.699 17.392]& [1399  1400] \\
                &7 & 10.159&5.386 &\\
                &  & [ 9.693 10.623] & [4.079 7.218]& \\
                \hline \hline
            \end{tabular}
            \caption{Simulated example -  second scheme - Model2:  posterior means  $ (\,\hat{ }\, )$  and credible intervals (CI) of  $\mu_k$, $\sigma_k^2$ and  $\xi_k$. } \label{tab:ex4}
        \end{table}

        We simulate datasets under two schemes, both with $T=1500$, 7 regimes and  assuming conditional  independence between the $y_t$'s and with  $Y_t|\boldsymbol{\theta}_k \sim N(\mu_k,\sigma_k^2)$. 
        In the first set  the change points are $\boldsymbol{\xi}=\{\xi_k\}=\{50, 250, 650$ $750, 1000, 1400 ,1500\}$, and the parameters are $\boldsymbol{\mu}$ $=\{\mu_k\}_{k=1}^7$ $=\{0, 5, 2,-2, 0, 2, 10\}$ and  $\boldsymbol{\sigma}^2=\{\sigma_k^2\}_{k=1}^7=\{1,2,1,$ $0.5,1,3,5\}$ while in the other $\boldsymbol{\xi}=\{50, 250, 900, 950,$ $ 1100, 1400 ,1500\}$ with $\boldsymbol{\mu}=\{0,5,2,2,2,2,10\}$ and  $\boldsymbol{\sigma}^2=\{1,2,1,0.1,1,15,5\}$. 
        For each scheme 100 datasets are simulated; two examples of simulated time series, one for each scheme, are  plotted in Figures \ref{fig:exy1} and \ref{fig:exy12}.

    {The set of  parameters of the first scheme are chosen so to have regimes of short (1 and 4) and long (3 and 6) length, overlapping   distributions on adjacent regimes (2-3 and 4-5-6), well separated ones (1-2 and 3-4) and different values of variability. In the second scheme we are mainly interested in the evaluation  of how the models behave   when a short regime (the forth) is in  between two regimes  (the third and fifth)  that have the same density parameters. }

        We assume a normal prior for $\mu_k$ with 0 mean and variance 1000, while  the prior over $\sigma_k^2$ is  inverse gamma with shape and rate parameters  both equal to 1. 
        Here $\sigma_{\beta}^2$ is set  to be 1000 and through a simulation  we evaluated that it induces a prior over $K$ that puts the central 90\% of probability mass between 741 and  1477.   For  Model2, we estimated model with $K^* \in\{ 4,5,\dots ,10 \}$.

        \paragraph{First scheme}
        
        Under our proposal the   maximum a posteriori (MAP) estimate of $K$ is 7 in 96 datasets, in 3 is 8 and in  1 is 9. We measure the agreement between the true partition and the one found by our model, i.e. the MAP classification,through  the  Rand Index (RI)  \citep{Hubert1985}, that is an index that  ranges between 0 and 1, with 0 indicating that the  partitions, the true one and the MAP,  do not agree in any pair of points and 1 in case of perfect agreement \citep[for details see][]{Hubert1985}. Among datasets, the minimum value reached by the RI is 0.986 and in 27 of them is exactly 1.   Model2 identifies 7 regimes in 99 datasets and 8 in 1, with minimum RI  equal to 0.969, that is   lower than the one found by our model, and it is exactly 1 in 27 datasets. On the other hand,
        Model3 identifies always 1 regime and  we will give a justification  of this result in the Appendix.

        For Model1 and Model2 we show in Tables \ref{tab:ex} and \ref{tab:ex2}  the posterior estimates and credible intervals (CIs) for the simulated dataset depicted in    Figure \ref{fig:exy1}.  Results under Model1 and Model2 are quite similar and both  estimate well the parameters, i.e. the true values are  inside the associated CIs, but 
        only with our proposal we have an estimate of the posterior distribution of $K$, shown in Figure \ref{fig:K1},  and we can evaluate the uncertainty on the estimated number of regimes.  The CI of $\beta$ is [0.076 0.366] for Model1 and [0.0730 0.352] for Model2.

        \paragraph{Second scheme}
        
        {Under Model1, the MAP estimate of $K$ is 7 in 75 datasets, while is 8 in 6, 9 in 1, 6 in 1 and 5 in 17. As in the first scheme, Model3 identifies always 1 regime while Model2 estimates 7 regimes in 23 datasets and 5 in 66. As we expected, when a posterior sample of $K$ is 5, regimes 3, 4 and 5 are generally merged.}\\
        In terms of RI  Model1 and Model2 have similar minimum and maximum values, i.e. for both the  maximum is 0.999 while Model1 has minimum 0.717 and Model2 0.714.  The main difference is in the distribution of the RI across simulated datasets, since our proposal has median RI that is equal to 0.995 while the model of \cite{Chib1998} has median value 0.725, showing that  our proposal tends to  perform better.

        For the simulated dataset plotted in Figure \ref{fig:exy12}, we show in Tables \ref{tab:ex3} and \ref{tab:ex4} the parameters estimates for Model1 and Model2, where we can see a good  agreement between the posterior CIs;  for both models the values $\sigma_{3}^2$ and $\sigma_{4}^2$ are the only one not inside the associated CIs.  $\beta$   has CI [0.0778 0.370] in Model1 and [0.076 0.369] under Model2.  The posterior distribution of $K$ is shown in Figure \ref{fig:K2}.

        \subsection{Coal-mining disasters}
        \begin{figure*}[t]
            \centering
            {{\includegraphics[scale=0.48]{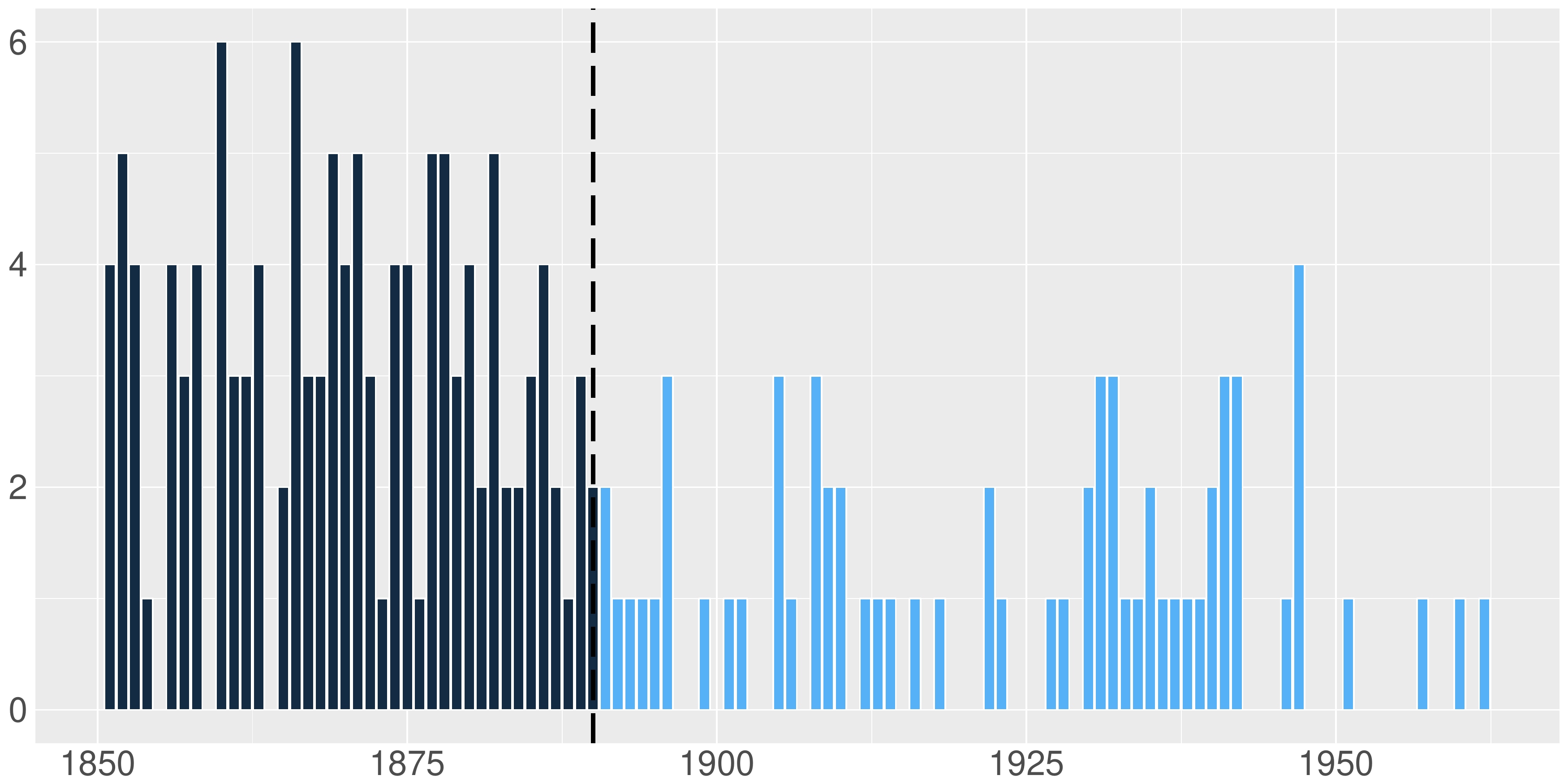}}}
            \caption{Coal-mining disasters data: the height of the bar represents the count in the associated year. The vertical dashed line separates the regimes identified by the MAP estimate of Model1.}\label{fig:exy2}
        \end{figure*}

        \begin{figure*}[t]
            \centering
            {\includegraphics[scale=0.45]{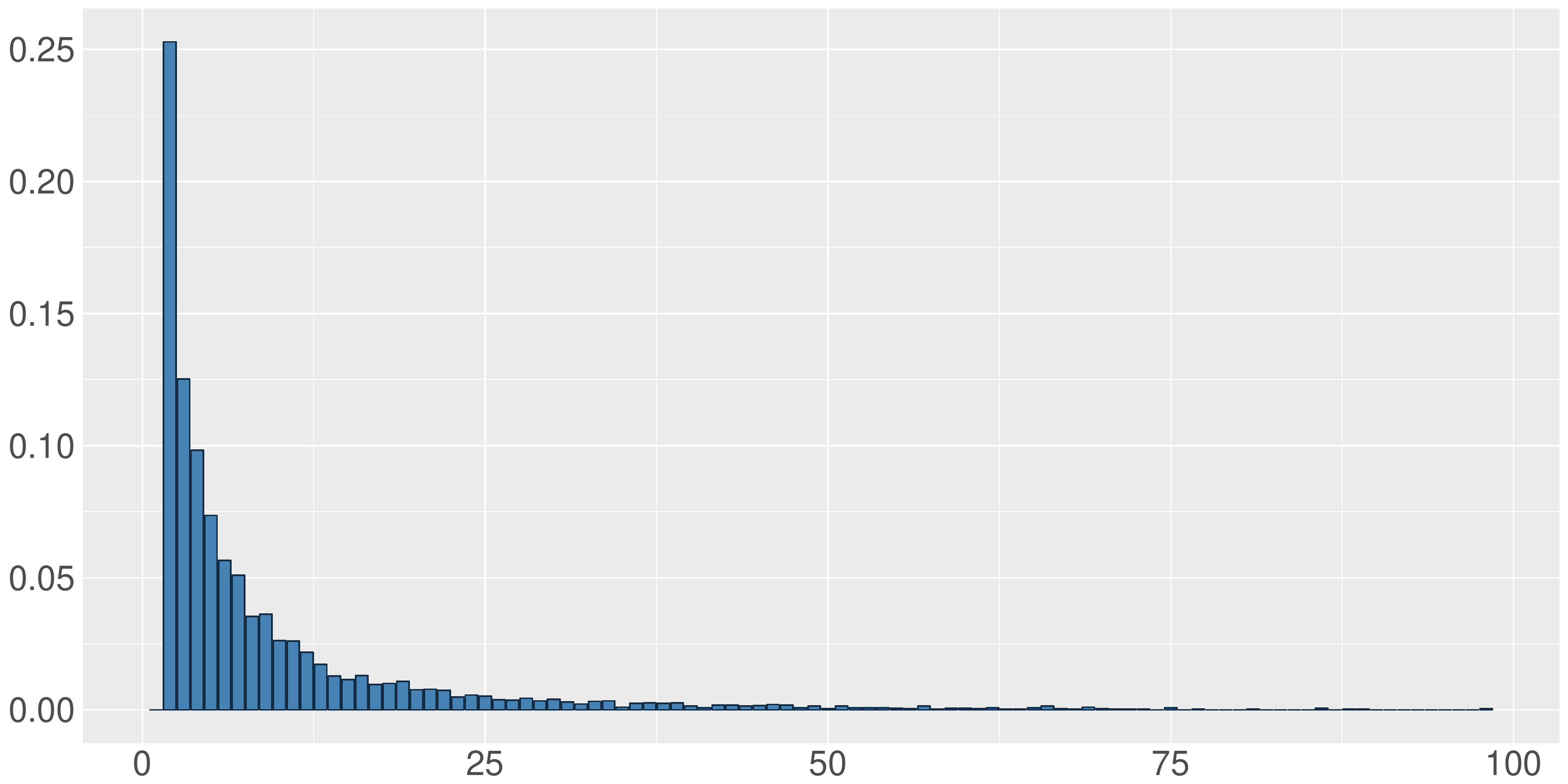}}
            \caption{Coal-mining disasters data: posterior distribution of $K$.}\label{fig:coalK}
        \end{figure*}

        Our first real data application  is devoted to the analysis of one of the most analysed dataset in change-point literature \citep[see][]{JARRETT1979,Carlin1992}; the annual number of coal-mining  disasters in  Britain, during the period 1851-1962. Here $y_t \in \mathbb{N}$, $\boldsymbol{\theta}_k = \lambda_k$  and, following \cite{Chib1998}, we assume  $f(y_{t}|\{ y_j \}_{j=1}^{t-1}, \lambda_{k})=f(y_{t}| \lambda_{k})$
        with $Y_t| \lambda_{k}\sim Pois(\lambda_{k})$.  The data  are plotted in Figure   \ref{fig:exy2}.

        We assume $\lambda_k\sim G(2,1)$  while the variance parameter of the half normal prior on $\beta$ is set to 0.1 and, again through a simulation, we evaluated that this induces a prior over $K$ that puts the central 90\% percent of total mass of probabilities between 1 and 10.  On this dataset the Model2 is tested with $K$ between 1 and 4.

        The MAP estimate of $K$ is 1  under Model3, i.e. there is not a segmentation, while Model2 chooses $K=2$ that is also the value found in  \cite{Chib1998}. Our proposal  has MAP estimate of $K$ equal to 2 and the associated posterior distribution is showed in Figure  \ref{fig:coalK}.  We wanted to point out that since our proposal has  a non-parametric specification of the latent allocation structure, it is not surprising that with little data, as in this example,  the posterior of $K$ has a lot of variability.

        For Model1, posterior mean estimates of $\lambda_{1}$ and $\lambda_{2}$  are   $\hat{\lambda}_1=3.045$ and $\hat{\lambda}_2=0.923$ with, respectively, CIs  [2.544 3.648]  and [0.711 1.166], on the other hand, under Model2 they are  $\hat{\lambda}_1=3.084$ and $\hat{\lambda}_2=0.933$ with, respectively, CIs  [2.587  3.688]  and [0.7223  1.186].  In both models the CI of $\xi_1$ is [1886 1896] with MAP estimate 1890 while the CI of $\beta$ is [0.053 1.017] for Model1 and [0.025 0.45] for Model2.

        \subsection{Indoor radon data} \label{sec:radon}

        \begin{figure*}[t]
            \centering
            {{\includegraphics[scale=0.46]{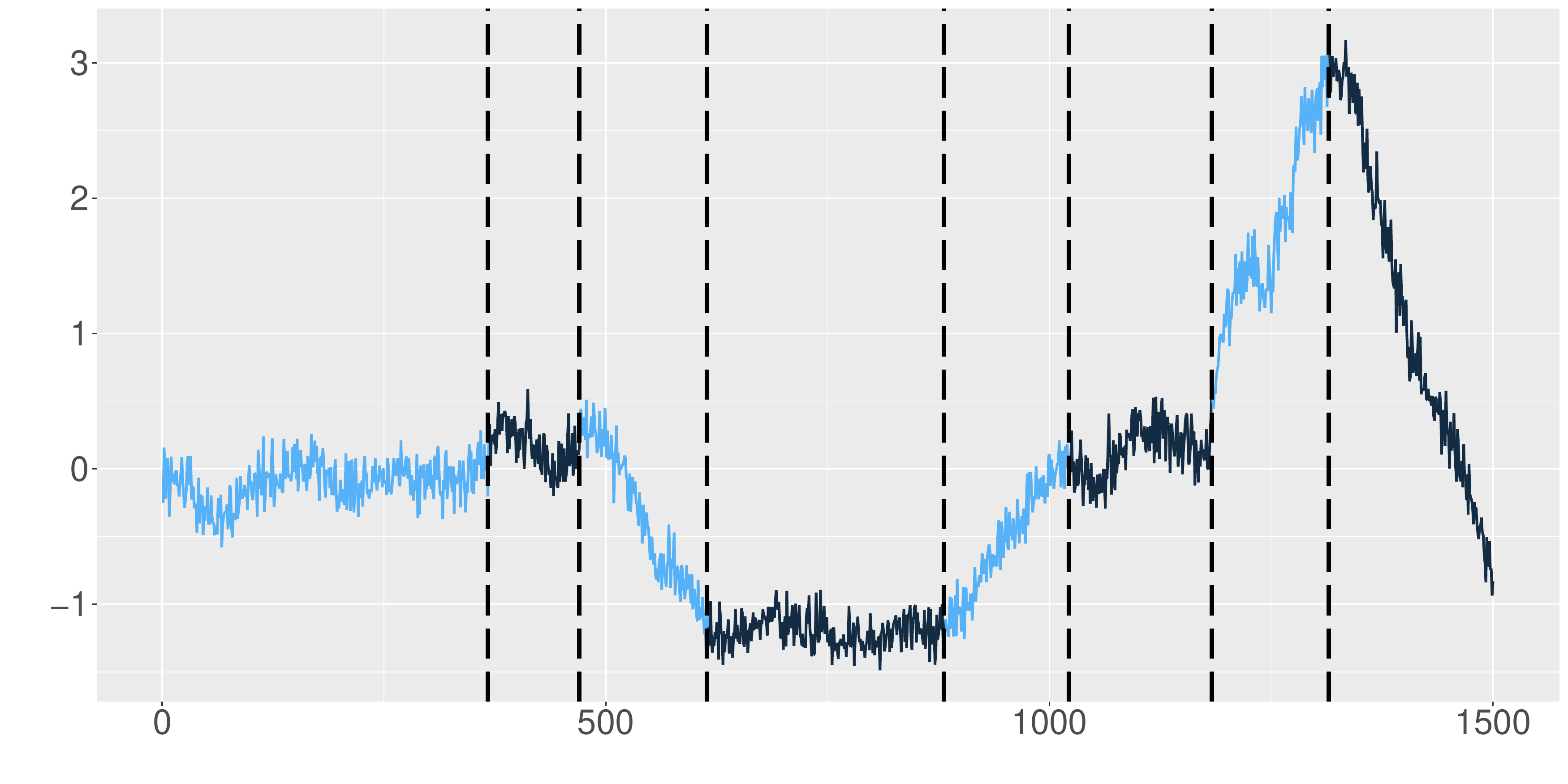}}}
            \caption{Indoor radon concentration data:   the vertical dashed line separates the regimes identified by the MAP estimate of Model1.}\label{fig:radon}
        \end{figure*}
        
        \begin{table*}[t]
            \centering
            \begin{tabular}{cc|ccccc}
                \hline \hline
                &&$\hat{\mu}_{0,k}$&$\hat{\mu}_{1,k}$& $\hat{\sigma}_{k}^2$ & $\hat{\xi}_k$\\
                && CI & CI& CI&CI\\
                \hline
                &1 & -0.19 &0  &0.027& 367\\
                & & [-0.225 -0.155] &  [0.000 0.001]&[0.024 0.032] &[366   371 ] & \\
                &2 & 1.429 &-0.003 &0.039& 470\\
                & & [0.856 1.987] &  [-0.004 -0.002]&[0.030 0.052] &[468   471] & \\
                &3 &  5.824&-0.011 &0.032&614 \\
                & & [5.448 6.218] &  [-0.012 -0.011]&[0.026 0.041] &[604   627] & \\
                &4 &-1.136  & 0&0.021&881 \\
                & & [-1.311 -0.953] &  [-1.311 -0.953]&[0.018 0.025] &[865   893] & \\
                Regime  &5 &-9.571  &0.01 &0.027&1022 \\
                & & [-10.360  -8.943] &  [0.009 0.010]&[0.022 0.035] &[1004  1027] & \\
                &6 & -2.042 & 0.002&0.041&1183 \\
                & & [-2.862 -1.243] &  [0.001 0.003]&[0.034 0.051] &[1181  1187] & \\
                &7 & -18.73 &0.016 &0.069&1315 \\
                & & [-20.302 -17.178] &  [0.015 0.018]&[0.055 0.089] &[1309  1325] & \\
                &8 & 31.758 & -0.022&0.046& \\
                & & [30.886 32.641] &  [-0.022 -0.021]&[0.037 0.056] & & \\
                \hline \hline
            \end{tabular}
            \caption{Indoor radon concentration data: posterior means   $ (\,\hat{ }\, )$    and credible intervals (CI) of $\mu_{0,k}$, $\mu_{1,k}$, $\sigma_k^2$ and $\xi_k$ computed using the subset of posterior samples that has $K = 8$. } \label{tab:radon}
        \end{table*}

        \begin{table*}[t]
            \centering
            \begin{tabular}{cc|ccccc}
                \hline \hline
                &&$\hat{\mu}_{0,k}$&$\hat{\mu}_{1,k}$& $\hat{\sigma}_{k}^2$ & $\hat{\xi}_k$\\
                && CI & CI& CI&CI\\
                \hline
                &1 & -0.758 & 0.448 &1.495& 1 \\
                & & [-44.558  44.088] &  [-42.921  44.201]&[0.262 39.986] &[1 2] & \\
                &2 & -0.23 &0.001 &0.03&477 \\
                & & [-0.263 -0.197] &  [0.001 0.001]&[0.027 0.034] &[469   521] & \\
                &3 & 5.94 & -0.012&0.033&613 \\
                & & [5.245 6.525] &  [-0.013 -0.010]&[0.026 0.046] &[604   629] & \\
                &4 &  -1.142& 0&0.021&883 \\
                & & [-1.318 -0.952] &  [-1.318 -0.952]&[0.018 0.025] &[865   895] & \\
                Regime  &5 &-9.61  &0.01 &0.028&1021 \\
                & & [-10.584  -8.911] &  [0.009 0.011]&[0.022 0.037] &[988  1027] & \\
                &6 &  -2.017& 0.002&0.041&1183 \\
                & & [-2.912 -1.253] &  [0.001 0.003]&[0.032 0.052] &[ 1180  1188] & \\
                &7 &-18.712  & 0.016&0.068&1317 \\
                & & [-20.190 -17.031] &  [0.015 0.018]&[0.053 0.087] &[1308  1334] & \\
                &8 & 31.776 &-0.022 &0.046& \\
                & & [30.865 32.631] &  [-0.022 -0.021]&[0.038 0.056] & & \\
                \hline \hline
            \end{tabular}
            \caption{Indoor radon concentration data. Posterior means   $ (\,\hat{ }\, )$    and credible intervals (CI) of $\mu_{0,k}$, $\mu_{1,k}$, $\sigma_k^2$ and $\xi_k$.  } \label{tab:radon2}
        \end{table*}

        %In this section we analyze a radon {concentration measurements} time series. 
        Radon is  a colorless and odorless   inert noble  gas generated by the radioactive decay of Radium (Ra226) in the decay chain of Uranium (U238) \citep{Hauksson1981}.
        A time series of radon concentration is characterized by  daily and annual periodic components with 
        about  daily changes in mean level, variance and temporal trend \citep{barbosa2010}. 
        Radon concentration is considered a proxy of geodynamic activity since many authors  \citep[see][and references therein]{monnini1997,Woith2015} proved that prior to a powerful  geodynamic event, such as an earthquake, a radon time series can show abrupt and out of ordinary  changes \citep{Steinitz2003,Kawada2007}; this connection between radon anomalies and   geodynamic events  makes {relevant the understanding of the radon time series dynamic. }

        In this example we use  the radon data   owned by the \emph{International Association for Research Seismic Precursors} (iAReSP) \citep{Nicol2015,Nicoli2016a}.  The iAReSP, with the  Tellus project \citep{TELLUS}, that consists of a network of radon recording stations, 
        aims to understand what happens to the radon concentration during the phase preceding an earthquake. At the present moment we have data only from one station {and on a limited time window. More precisely
            our data are the  mean radon counts over ten minutes,  observed between November 18$^{th}$ 2015, 8:00, and  November 28$^{th}$ 2015, 17:50,  having then 1500 observations.}
        Data are recorded in central Italy, in the town of Pizzoli, close to the city of L'aquila, 803 $m$ above sea level, using an ionization chamber with continuous measurement of Alfa particles produced by the decay of radon stable isotope  $^{222}$RN  \citep{Nicol2015}. In the observational period no major earthquakes were observed.

        Here we show some  {preliminary} results that  prove the ability/potentiality of the model in the segmentation of a radon time series.  As said in the beginning of this section, radon data presents a daily periodicity that  is  {stable in time \citep[see][]{barbosa2010};  in other words changes in the time series do not affect   its amplitude. This characteristic of the radon data cannot be expressed in our model which   assumes that all parameters  change between regimes  and then, prior to the model fitting, we decompose the time series into seasonal, trend and irregular components using the approach of \cite{cleveland90}, implemented in the \emph{stl} function of \emph{R} and the (daily) seasonal component is subtracted  to eliminate the periodicity. }
        The resulting time series has mean  $\approx$5080.515 and variance 
        $\approx$1124047 and, to avoid possible numerical stability problems  that  such large numbers may raise,
        we  standardize the data;
        the resulting   time series is plotted  in Figure \ref{fig:radon}. 
        To take into account changes in mean level, variance and temporal trend the following is assumed:
        $Y_t| \boldsymbol{\theta}_k \sim N(\mu_{0,k}+\mu_{1,k} t, \sigma_{k}^2) $. \\
        Parameters $\mu_{0,k}$ and $\mu_{1,k}$ have normal  priors with 0 mean and variance 1000 while $\sigma_k^2\sim IG(1,1)$. In this example $T$ is 1500, as in the simulated ones, and then  we use the same prior on $\beta$.

        \emph{A posteriori}, Model3 estimates only 1 regime while our proposal put 99\% of the mass of probability on $K=8$ and the remaining on $K=9$, the posterior means and CIs of parameters and change points can be seen in Table \ref{tab:radon} while Figure \ref{fig:radon} shows the MAP classification. Model2 estimates 8 regimes. The main differences between our proposal and Model2 can be seen in the estimates of parameters and change points of the first two regimes, see Tables  \ref{tab:radon} and \ref{tab:radon2}.
        Posterior mean estimates, CIs and change points of the other regimes, are  similar. The CI of $\beta$ is [0.078 0.331] under Model1 and [0.082 0.368] under Model2.

        Model1 and Model2 found a clear and reasonable segmentation of the data showing that there  are almost daily changes in the radon emission features. This last finding is coherent with previous studies, see for example \cite{barbosa2010}.

        \section{Discussion} \label{sec:disc}

        In this work we proposed a semi-parametric formalization of   the standard change-point model  of \cite{Chib1998}. In our extension, the first order latent Markov process, ruled by a constrained one-step  transition matrix,   is substituted by a  stochastic process  based on the   stick-breaking representation of the DP. We suggested to draw samples from the posterior distribution using a marginalized version of the proposed model and we showed how to compute the full conditionals needed to implement the MCMC algorithm.  \\
        To asses the ability of the model in recovering the number and locations of change points, we used  simulated examples. Then we make inference 
        in one of the most analysed dataset in the literature and on a new one. We  showed  that  our proposal outperformed the ones of 
        \cite{ko2015} and \cite{Chib1998} in terms of change points estimates.\\  
        {The future will find us enriching the model including covariates information and modelling  multiple time series  subjected to individual and   concurrent shifting  in their features. We will also use the model to analyzed a longer times series of radon data to possibly acquire an early signal of major earthquake events.
        }
        
%       \section*{Online supplementary materials}
        
%       \begin{description}
%           \item[ReadMe:] Instructions on the usage of the codes (ReadMe.txt).
%           \item[Simulated examples:] Codes to perform the analyses on the simulated examples (/Simulated Datasets, folder).
%           \item[Coal-mining disasters:] Codes to perform the analyses on the coal-mining disasters data (/CoalMiningDisasters, folder).
%           \item[Functions:] Functions that implement the MCMC algorithms  (/Functions, folder).
%       \end{description}

        \section*{Acknowledgements}
        The author wishes to thank Giovanna Jona Lasinio and Antonello Maruotti for assistance and comments that greatly improved the manuscript.
        
        %%% ** The bibliograhy **
        \bibliographystyle{natbib}
        \bibliography{/Users/gianlucamastrantonio/Dropbox/tex/BibAll}% place <bib-data-file> 
        %\bibliography{/Users/gianlucamastrantonio/Desktop/statistico/paper/all}
        \appendix

        \section*{Appendix} \label{eq.fdsec:ko}
        
        Here we discuss  the problematic aspects  and  errors in the computation of the  full conditionals  in \cite{ko2015} that, in our opinion, justify the need of a more rigorous formalized model as the one we are proposing in this work.

        %\cite{ko2015} follow the work of \cite{Beal2002} that  does not formalize rigorously the associated DP. But then \cite{Teh2006} and \cite{teh2010} introduced the hierarchical Dirichlet process and showed how the model of \cite{Beal2002} can be formalized rigorously.
        
        % 
        % 
        %Moreover they define the DP as the limit for $k \rightarrow \infty$ of
        %\begin{equation}
        %\mathbf{p}_{i} \sim Dir\left( 0,\dots,0, \frac{\beta}{k-i+1}, \dots, \frac{\beta}{k-i+1} \right),
        %\end{equation}
        %(see equation (5) of  \cite{ko2015}) that implies a Dirichlet distribution for a vector with some of its element constant and equal to 0  and then deterministic. 
        %\paragraph{The model}
        
        To obtain e semi-parametric extension of \cite{Chib1998},
        \cite{ko2015} substitute the matrix $P$ with 
        \begin{equation}
        \tilde{P} = \left(
        \begin{array}{ccccccc}
        \pi_{11} & \pi_{12} & \pi_{13} & \dots & \pi_{1K^*}  \\
        0 & \pi_{22} & \pi_{23} & \dots & \pi_{2K^*}  \\
        0 & 0 & \pi_{33} & \dots & \pi_{3K^*}  \\
        \vdots & \vdots & \vdots & \vdots & \vdots \\
        0 & 0 & 0 & \dots & \pi_{K^*K^*}  \\
        \end{array}\right) ,\label{eq:p2}
        \end{equation} 
        and they assume  
        \begin{equation}
        [\tilde{P}]_k | \beta \sim Dir\left(  \frac{\beta}{K^*-k+1}, \dots, \frac{\beta}{K^*-k+1} \right) .\label{eq:pif}
        \end{equation}
        They integrate out   the elements of $\tilde{P}$ and  taking  the limit as $K^*$ goes to infinity,   the following time dynamic for $s_t$ is obtained:
        \begin{equation}
        f(s_{t}=i|s_{t-1}=k,s_1,\dots , s_{t-1},\beta)=
        \begin{cases}
        \frac{n_{k}^{1:(t-1)}}{n_{k}^{1:(t-1)}+\beta} & \mbox{if } i=k,\\ 
        \frac{\beta}{n_{k}^{1:(t-1)}+\beta} & \mbox{if } 
        %j \neq s_h,h=1,2,\dots , t
        i=k+1.\\
        \end{cases}
        \label{eq:s_t22}
        \end{equation}
        \cite{ko2015} noted that, 
        if $s_t=k$ and $s_{t+1}=k+1$, then $s_{t+2}$ will jump with probability 1 to a new state. To avoid the problem   a  self-transition  mass $\alpha \in \mathbb{R}^+$  is introduced and   \eqref{eq:s_t22} is modified as

        \begin{align}
        &f(s_{t}=i|s_{t-1}=k,s_1,\dots , s_{t-1},\beta, \alpha)=\\
    &\phantom{s}    \begin{cases}
        \frac{n_{k}^{1:(t-1)}+\alpha}{n_{k}^{1:(t-1)}+\alpha+\beta} & \mbox{if } i=k,\\ 
        \frac{\beta}{n_{k}^{1:(t-1)}+\alpha+\beta} & \mbox{if } 
        %j \neq s_h,h=1,2,\dots , t
        i=k+1.\\
        \end{cases}
        \label{eq:s_t2}
        \end{align}

        \paragraph{The model  formalization}
        
        The authors state that, as $K^*\rightarrow \infty$, each row of $\tilde{P}$ follows a DP. Since the DPs are independent it is not clear how   the  rows share the same set of DP atoms since there must be a way to tie them as in the hierarchical Dirichlet process of \cite{Teh2006}, i.e. the atom associated to  $\pi_{ij}$, with $j \geq i$, must be equal to the one of $\pi_{hj}$, for all  $j\geq h$. 
        This was also the problem of the  infinite hidden Markov model of \cite{Beal2002}, that is close to the proposal of \cite{ko2015}, but then the work of  \cite{Teh2006} shows how to solve this problem.
        % and I think that it can be also used   in the framework of \cite{ko2015}.
        %The very fact that the authors,  to avoid a computational problem,  introduce a new parameter $\alpha$,   it is probably due to this lack of rigorous formalization. %and, for example, if $f(s_1=1,s_2=1,s_3=2,s_4=2| \tilde{P})= \pi_{11}\pi_{12}\pi_{22}>0$ (see \eqref{eq:p2}) we wonder how it is possible that  (with a slight abuse of notation)
        %\begin{align}
        %ss
        %\end{align} 
        %$$
        %f(s_1=1,s_2=1,s_3=2,s_4=2|\beta) = \lim_{K^* \to \infty}  \int_{ \tilde{P}} \pi_{11}\pi_{12}\pi_{22} f( \tilde{P} | \beta) d  \tilde{P} =0
        %$$
        %since, as we can see from equation \eqref{eq:s_t22}, $f(s_4=2|s_1=1,s_2=1,s_3=2,\beta)=0$.
        %

        \paragraph{The joint density of $\{s_1,\dots,s_T|\beta\}$}

        Equation \eqref{eq:s_t2} reduces to our specification if $\alpha=1$, see equation  \eqref{eq:s_t}.  Then the joint density  of  $s_1,\dots,s_T|\alpha=1,\beta$ derived from \cite{ko2015} and  $s_1,\dots,s_T|\beta$ derived in this work must be  the same. \cite{ko2015} write
        \begin{equation}\label{eq:ab2}
        f(s_1,\dots,s_T|\alpha=1,\beta)= \beta^{K} \prod_{i=1}^{K}\frac{\Gamma(\alpha+\beta)}{\Gamma(\alpha)} \frac{\Gamma(n_{i}^{1:T}+\alpha  )}{\Gamma(n_{i}^{1:T}+1+\alpha+\beta  )}.
        \end{equation}
        \eqref{eq:ab2}  is different from our   \eqref{eq:S}. 
        Equation \eqref{eq:ab2} implicitly  assumes the existence of  a new observation at time $T+1$  belonging to the $(K+1)^{th}$ regime, and  if we 
        multiply \eqref{eq:S} by $f(s_{T+1}=K+1|s_{T}=K,s_1,\dots , s_{T-1})=\frac{\beta}{n_{K}^{1:T}+1+\alpha+\beta}$, we obtain   \eqref{eq:ab2}.

        \paragraph{The full conditional of $s_t$}

        When \cite{ko2015} show how to sample from the full conditional of $s_t$, they  derive the following probabilities: 
        \begin{align}
    &   f(s_{t+1}=k+1|s_t=k,s_{t+2},\dots s_{T},\alpha,\beta)  =\\ & \phantom{s} \frac{\beta}{n_{k}^{1:(t-1)}+1+\beta+\alpha}\label{ko1},\\
    &   f(s_{t+1}=k+1|s_t=k+1,s_{t+2},\dots s_{T},\alpha,\beta) =\\& \phantom{s}  \frac{n_{k+1}^{(t+1):T}+\alpha}{n_{k+1}^{(t+1):T}+\beta+\alpha} \label{ko2}.
        \end{align}
        Just for example let assume $s_t=s_{t-1}=s_{t-2}=s_{t-3}=k$ and  $s_{t-4}=k-1$; then  $n_{i}^{1:(t-1)}=2$.  Indeed we can find  $n_{k}^{1:(t-1)}$ only if we know $\{s_{t-3},s_{t-2},s_{t-1}\}$ and then it is not  possible that $n_{k}^{1:(t-1)}$ appears  in equation \eqref{ko1} if none of them  is in the conditioning set of \eqref{ko1}.
        As further example, let assume  $s_{t+2}=i+1$. In this case it is trivial to demonstrate that $f(s_{t+1}=k+1|s_t=k+1,s_{t+2}=k+1,\dots s_{T},\alpha,\beta) =1$ and not $\frac{n_{k+1}^{(t+1):T}+\alpha}{n_{k+1}^{(t+1):T}+\beta+\alpha}$, as in equation  \eqref{ko2}.

        \paragraph{The MCMC algorithm}
        
        In the  MCMC algorithm proposed in   \cite{ko2015},  during the model fitting the number of occupied regimes  can only decrease or   remain the same. 
        The major implication  is  the possibility that the MCMC can  never reach the stationary distribution and then we cannot   sample from the posterior. To see more clearly why, let suppose that the posterior distribution of $K$, the number of change points, is entirely concentrated on $d$, i.e. $f(K=d|\{y_t \}_{t=1}^T)=1$, and we initialize the MCMC with $K+c$ different regimes ($c \in \mathbb{N}$). It can happen that at the $b^{th}$ iteration, before the chain has reached its stationary distribution,  $K$ assumes value $d-1$. Then, since $K$ can only decrease or remain constant, after the $b^{th}$ iteration, it will never assume value $d$, the chain will never reach its stationary distribution and we cannot have samples from the posterior  of interest.  Notice that, given samples from the MCMC algorithm,  we have no way to tell if they are from the posterior distribution or not \citep{brooks2011}. \\
        Moreover, even if the algorithm has reached its stationary distribution, there is always a non zero probability   that two regimes can be merged (see  the results of the  second scheme in Section \ref{sec:simex}) and then, if we run the algorithm for enough iterations, eventually this  will happen and the number of change points decreases with no possibility to increase again. As a consequence, as the iterations  go toward infinity, the number of regimes  tends to  1. This last consideration   explains why in the  examples of Section \ref{sec:ex}, the algorithm of \cite{ko2015}  finds always 1 regime.

%\begin{thebibliography}{}
%%
%% and use \bibitem to create references. Consult the Instructions
%% for authors for reference list style.
%\end{thebibliography}

\end{document}